\documentclass[twocolumn]{aastex62}
\usepackage{amsmath,amstext}
\usepackage[T1]{fontenc}
\usepackage{natbib}
\usepackage{sidecap}

\newcommand\psj{Planetary Science Journal}
\begin{document}

\title{Orbital dynamics landscape near the most distant known trans-Neptunian objects} 
\shorttitle{}

\author[0000-0001-8736-236X]{Kathryn Volk}
\affiliation{Lunar and Planetary Laboratory, The University of Arizona, 1629 E University Blvd, Tucson, AZ 85721}
\correspondingauthor{Kathryn Volk}
\email{kvolk@lpl.arizona.edu}

\author[0000-0002-1226-3305]{Renu Malhotra}
\affil{Lunar and Planetary Laboratory, The University of Arizona, 1629 E University Blvd, Tucson, AZ 85721}

\begin{abstract}

The most distant known trans-Neptunian objects (perihelion distance above 38 au and semimajor axis above 150 au) are of interest for their potential to reveal past, external, or present but unseen perturbers. 
Realizing this potential requires understanding how the known planets influence their orbital dynamics. 
We use a recently-developed Poincare mapping approach for orbital phase space studies of the circular planar restricted three body problem, which we have extended to the case of the three-dimensional restricted problem with $N$ planetary perturbers. 
With this approach, we explore the dynamical landscape of the 23 most distant TNOs under the perturbations of the known giant planets. 
We find that, counter to common expectations, almost none of these TNOs are far removed from Neptune's resonances.
Nearly half (11) of these TNOs have orbits consistent with stable libration in Neptune's resonances; in particular, the orbits of TNOs 148209 and 474640 overlap with Neptune's 20:1 and 36:1 resonances, respectively. 
Five objects can be ruled currently non-resonant, despite their large orbital uncertainties, because our mapping approach determines the resonance boundaries in angular phase space in addition to semimajor axis. 
Only three objects are in orbital regions not appreciably affected by resonances: Sedna, 2012 VP113 and 2015 KG163. 
Our analysis also demonstrates that Neptune's resonances impart a modest (few percent) non-uniformity in the longitude of perihelion distribution of the currently observable distant TNOs.  
While not large enough to explain the observed clustering, this small dynamical sculpting of the perihelion longitudes could become relevant for future, larger TNO datasets.

\end{abstract}


\section{Introduction} \label{sec:intro}

There has been much recent discussion about the orbits of the most distant set of observed trans-Neptunian objects (TNOs) and whether or not they offer evidence for an as-yet-undetected outer solar system planet \citep[e.g.,][]{Trujillo:2014,Batygin:2016,Malhotra:2016,Madigan:2016,Shankman:2017,Shankman:2017o,Napier:2021,Brown:2021, Bernardinelli:2022}.
The orbits of TNOs that are dynamically `detached' from strong interactions with Neptune and the other giant planets are of particular interest in looking for signs of undiscovered distant perturbers.
The orbital distribution of the detached TNOs could also record long-gone perturbers such as rogue planets ejected from the early solar system \citep[e.g.][]{Gladman:2006}, past strong external perturbations such as close encounters between stars and the solar system \citep[e.g.][]{Kaib:2011,Brasser:2012}, or details such as the `graininess' of Neptune's migration to its current orbit \citep[e.g.][]{Kaib:2016b,Lawler:2019}.
However, identifying the `detached' objects is fraught with ambiguities.
Many of the observed TNOs with very long orbital periods, i.e., with large semimajor axes $a$, belong to the scattering population (see, e.g., \citealt{Gladman:2008} for commonly used definitions of TNO dynamical classes). 
Most of these scattering objects have perihelion distances $q\lesssim37-38$~au, low enough to cause relatively strong gravitational perturbations from Neptune.
In general, the larger the value of $a$, the larger the value of $q$ below which planetary perturbations induce a significant random walk in $a$ over the age of the solar system \citep[see, e.g., discussion in][]{Bannister:2017,Gladman:2021}.
TNOs with larger $q$ that do not experience significant changes in $a$ over time due to the known giant planets are more likely to preserve signatures of other perturbers.
However, the potential for mean motion resonances (MMRs) with Neptune out to high $a$ and high $q$ complicates whether an object can truly be considered as detached.
Secular resonances within these MMRs can induce long-term oscillations in eccentricity that could bring an object close enough to experience scattering at some future epoch even if its current perihelion distance appears out of reach \citep[see, e.g.,][]{Gomes:2005}.

In previous work, different authors have used different cuts in $a$ and $q$ ($a\gtrsim150-250$~au and $q\gtrsim30-45$~au) in an attempt to isolate TNOs that are not strongly dynamically influenced by the giant planets, but these cuts are often not robustly motivated and thus fairly arbitrary.
To determine which of these TNOs are `detached' from the giant planets, some works have taken the approach of integrating the orbits of objects  meeting these rough cuts in $a$ and $q$ over long timescales (up to several Gyr; e.g., \citealt{Batygin:2019}) to look for mobility in semimajor axis, keeping only objects with relatively stable orbits.
However these analyses do not fully account for orbital semimajor axis uncertainties,
which can be as large as several au even for objects observed over many oppositions,
and do not adequately assess the potential for resonant interactions with Neptune.
Classification schemes that do adequately search for resonant behavior amongst potentially detached objects \citep[e.g.][]{Gladman:2008} are often limited to shorter integration timescales ($\sim10$Myr) and a sparse sampling of the orbital uncertainties. 

Our goal in this paper is to develop a more rigorous method to identify detached objects in the observed TNO population.
Our approach is based on a recently developed implementation of Poincar\'e return maps of the planar circular restricted three body model \citep{Wang:2017,Lan:2019,Malhotra:2021}. 
We extend this method to the restricted case of the three-dimensional N-body model to explore the dynamical landscape near the most distant observed TNOs under the influence of the known giant planets.
Understanding the dynamical landscape around the observed objects allows us to understand the likelihood of different dynamical states even for TNOs with large or poorly characterized orbital uncertainties. 
The rest of this paper is organized as follows.
In Section~\ref{s:obs}, we list the orbital data of the known TNOs under consideration and introduce the motivation for our approach with Poincar\'e return maps rather than direct integration of the (quite uncertain) observed orbits.
In Section~\ref{s:sims}, we describe the methodology of computing the Poincar\'e return maps to explore the dynamical neighborhood of each of the observed TNOs (expanded upon in Appendix~\ref{appendix}) and our resulting determination of the role (or lack thereof) of Neptune's distant resonances in their long term evolution. 
In Section~\ref{s:summary}, we discuss the implications of these results for future studies of the distant solar system.

\section{Orbital data of observed distant objects}\label{s:obs}

We use three main sources of orbital data for TNOs, the database at the Minor Planet Center (MPC)\footnote{\url{https://www.minorplanetcenter.net/db_search}}, the JPL small body browser \footnote{\url{https://ssd.jpl.nasa.gov/tools/sbdb_lookup.html}}, and JPL Horizons\footnote{\url{https://ssd.jpl.nasa.gov/horizons/app.html}}. 
To construct the list of TNOs we consider in this paper, we started with the list of all objects in the MPC database as of October 12, 2021 that meet the following criteria:
heliocentric perihelion distance $q>38$~au and semimajor axis $a>150$~au.
We then use the JPL small body browser to limit the list to only those objects with semimajor axis uncertainties $\delta a<10$~au.
Finally, we then use JPL Horizons to determine the TNOs' barycentric orbital elements, keeping only objects with barycentric perihelion distance $q>38$~au and semimajor axis $a>150$~au (for outer solar system objects, barycentric elements are much more dynamically relevant than heliocentric ones; see, e.g., \citealt{Gladman:2008,Gladman:2021}).
We choose these cutoffs in order to obtain a generous and inclusive sampling of the objects that have been considered `extreme' in the literature. 
The $q>38$~au boundary excludes most objects experiencing strong scattering from Neptune on megayear timescales. (However, as we will see, it does not exclude objects that may have moderate mobility in orbital energy/semi-major axis on long timescales.)
Our lower boundary in $a$ matches some of the lower-$a$ cuts in the literature \citep[e.g.][]{Trujillo:2014,Malhotra:2016,Bernardinelli:2020} to ensure our list of TNOs is inclusive.
The resulting list of TNOs are given in Table~\ref{t:objs}, where we list their primary designation, their barycentric orbital elements at the time of perihelion, and JPL's estimate of the 1-$\sigma$ uncertainty in each TNO's heliocentric semimajor axis (none of the available databases give both a barycentric orbit fit and an uncertainty, forcing us to mix coordinate systems; orbital uncertainties are discussed further below).

\begin{deluxetable*}{|l|r|r|r|r|r|r|r|l|r|}[htpb]
\tablecaption{The sample of TNOs investigated in this work.}
    \tablehead{object & $a$ & $\delta a$ & $q$ & $e$ & $i$ & $\omega$ & $\Omega$ & $T_p$ & $\psi$ \\
    designation & (au) & (au) & (au) & ($\circ$) & ($\circ$) & ($\circ$) & ($\circ$) & (JD) & ($\circ$)} 
    \startdata
   2015 KG163  &    679.8  &    5.1    &   40.49   &   0.9404  &    13.99  &     32.1  &    219.1  & 2459754.81 & 256.9	\\
 90377  Sedna$^\bullet$  &    506.4  &    0.17   &   76.19   &   0.8495  &    11.93  &    311.3  &    144.4  & 2479347.99	& 345.2 \\
   2013 RA109$^\bullet$  &    463.0  &    2.1    &   46.01   &   0.9006  &    12.40  &    262.9  &    104.8  & 2454266.49	& 46.7 \\
   2015 RX245$^\bullet$  &    423.7  &    5.2    &   45.55   &   0.8925  &    12.14  &     65.1  &      8.6  & 2475608.68 & 345.4  \\
   2016 SD106* &    350.2  &    3.8    &   42.70   &   0.8781  &     4.81  &    162.6  &    219.4  & 2464562.83 & 359.6     \\
   2010 GB174$^\bullet$  &    348.7  &    7.3    &   48.59   &   0.8606  &    21.56  &    347.3  &    130.7  & 2433928.47 & 279.2  \\
      474640$^\bullet$   &    327.7  &    1.7    &   47.32   &   0.8556  &    25.55  &    327.0  &     66.0  & 2455064.31	& 67.4 \\
   2013 SL102  &    314.5  &    0.75   &   38.13   &   0.8788  &     6.50  &    265.5  &     94.7  & 2455323.11	& 32.8 \\
    2015 GT50  &    311.2  &    2.5    &   38.41   &   0.8765  &     8.79  &    129.0  &     46.1  & 2451595.57	& 229.8 \\
    2013 FT28$^\bullet$  &    291.7  &    1.6    &   43.50   &   0.8509  &    17.38  &     40.7  &    217.7  & 2473469.28	& 182.1 \\
   2014 WB556  &    280.4  &    1.1    &   42.70   &   0.8477  &    24.16  &    235.3  &    114.9  & 2451283.61 & 46.8 \\
   2012 VP113$^\bullet$  &    261.9  &    1.5    &   80.52   &   0.6926  &    24.05  &    293.9  &     90.8  & 2443932.38	& 126.4 \\
   2016 SA59*  &    245.0  &    0.6    &   39.09   &   0.8456  &    21.50  &    200.2  &    174.6  & 2453106.10 &  60.8     \\
    148209$^\bullet$    &    222.0  &    0.6    &   44.12   &   0.8012  &    22.76  &    316.7  &    128.3  & 2438857.48	& 216.6 \\
      505478$^\bullet$   &    200.2  &    0.8    &   43.92   &   0.7806  &    10.65  &    252.1  &    192.0  & 2476000.65	& 353.4\\
  2003 SS422   &    190.7  &    0.7    &   39.59   &   0.7924  &    16.80  &    206.8  &    151.1  & 2454394.39 & 35.9 \\
  2015 UN105*  &    185.0  &    2.2    &   41.41   &   0.7761  &    37.02  &    231.5  &    129.4  & 2453134.01 & 46.4 \\
    2016 QV89  &   171.62  &    0.08   &   39.95   &   0.7672  &    21.39  &    281.1  &    173.2  & 2469915.13	& 39.8 \\
    2018 AD39  &    165.8  &    7.6    &   38.67   &   0.7668  &    19.77  &    49.22  &    330.1  & 2428376.33 & 213.1  \\
      506479   &    159.6  &    0.35   &   38.10   &   0.7613  &    15.50  &     10.8  &    197.9  & 2454873.70 & 243.6	\\
    2005 RH52  &    153.7  &    0.18   &   39.00   &   0.7463  &    20.45  &     32.5  &    306.1  & 2452802.96	& 26.1\\
   2015 KH163  &    153.0  &    0.58   &   39.94   &   0.7390  &    27.14  &    230.8  &     67.6  & 2471713.22	& 233.0 \\
   2013 GP136  &    150.2  &    0.19   &   41.04   &   0.7268  &    33.54  &     42.6  &    210.7  & 2465015.57	& 227.6 
       \enddata
    \tablecomments{The listed orbital elements are barycentric elements in the J2000 reference frame at the time of perihelion passage ($T_p$; last column) as given by JPL Horizons (\url{https://ssd.jpl.nasa.gov/horizons.cgi}); $\delta a$ is the 1$\sigma$ uncertainty in the heliocentric $a$ from the JPL small bodies database browser (\url{https://ssd.jpl.nasa.gov/sbdb.cgi}). Data was retrieved August 30, 2021 for all objects except the three denoted with *, which were retrieved October 12, 2021. Objects denoted with $\bullet$ are those that overlap with the TNO sample considered by \cite{Brown:2021} for their constraints on the unseen Planet Nine (see Section~\ref{s:discussion}).
    }
    \label{t:objs}
\end{deluxetable*}

A significant source of the uncertainty in an observed TNO's orbit is the fact that their orbital periods are extremely long compared to the timespan over which they have been observed ($\sim5-10$ years for most of the objects in Table~\ref{t:objs}). 
Determining an object's semimajor axis requires determining its orbital energy, including its kinetic energy, i.e., velocity.
A TNO's long orbital period means that the line-of-sight between the observer and the TNO changes slowly, which means that the observational constraint on the line-of-sight component of the velocity takes longer to constrain than for inner solar system objects, especially if the TNO is only observed near opposition (see discussion in, e.g., \citealt{BK}). 
In addition to uncertainties due to observational arc length, the uncertainties in the observationally measured astrometric positions contribute to the orbit-fit uncertainties.
Random errors in the astrometry are, in principle, relatively easy to quantify and propagate; however these errors are not reported in the astrometry databases and so must be estimated.
Moreover, there are also systematic errors in the astrometry (such as stellar catalog errors) that are more difficult to quantify and account for (see, e.g., discussion in \citealt{Gladman:2008}).
Different orbit fitting techniques also yield different orbital uncertainty estimates.
Thus it is very difficult to provide a precise accounting for the orbit-fit uncertainties of any given TNO.

We examine the orbit of TNO 2013 SL102 to show a typical example of the range of estimated orbit-fit uncertainties for a single TNO.
In Table~\ref{t:sl102}, we give orbit fits and uncertainties for 2013 SL102 from two commonly used orbit databases and two orbit fitting software packages.
We list best-fit $a$ values and 1-$\sigma$ uncertainties from the JPL small body database browser and the AstDys-2 database\footnote{\url{https://newton.spacedys.com/astdys/}} as well as from the \cite{BK} orbit fitting software package; for comparison to each of these orbit fits, we also used the online Find\_Orb orbit-fitting software\footnote{\url{https://www.projectpluto.com/fo.htm}} to produce an orbit-fit for the same epoch in the same coordinate system.
All the orbits in Table~\ref{t:sl102} were fit using the same 29 astrometric measurements from observations of 2013 SL102 (taken from the MPC) spanning a 3.2 year arc.
Note that the large change in heliocentric $a$ between the JPL and AstDys-2 fits is due to Jupiter's influence on the Sun's position and velocity in the two different epochs; barycentric $a$ values (such as given by the \citealt{BK} software and barycentric fits from Find\_Orb) do not suffer from this volatility.
The 1-$\sigma$ uncertainty estimates for $a$ range from 0.23-1~au ($\delta a/a = (0.7-3)\times10^{-3}$), a factor of $\sim4$; note also that the best-fit value of $a$ can differ by up to 50\% of the 1-$\sigma$ uncertainty between different orbit fits for the same epoch.
This illustrates why we prefer to take a mapping approach (described in more detail in the next section) in this work, rather than rely on any specific orbit fit and uncertainty estimate to determine the range of orbital phase space to investigate.
In addition to illustrating the dynamical regime near an observed orbit, including extending past the current estimated orbital uncertainties, these maps will also be useful for reference against future improved orbit fits.

\begin{deluxetable}{|l|l|l|l|l|l|}
\tablecaption{An example range of best-fit semimajor axis and uncertainties for 2013 SL102}
    \tablehead{source & $a$ & $\delta a$ &  $\delta a/a$ & coordinate & epoch \\
     & (au) & (au) & $\times10^{-3}$ & system & (JD)} 
    \startdata
    JPL & 337.88 & 0.75 & 2.2 & heliocentric & 2459396.5 \\
    Find\_Orb & 337.75 & 0.51 & 1.5 & heliocentric  & 2459396.5 \\ 
    \hline 
    AstDys-2 & 332.36 & 1.01 & 3.0 & heliocentric & 2459200.5 \\
    Find\_Orb & 332.24 & 0.49 & 1.5 & heliocentric  & 2459200.5 \\ 
        \hline 
    Find\_Orb & 314.33 & 0.44 & 1.4 & barycentric  & 2456563.6 \\     
       \enddata
    \tablecomments{The best-fit and 1-$\sigma$ uncertainties in 2013 SL102's semimajor axis take from JPL (\url{https://ssd.jpl.nasa.gov/sbdb.cgi}), AstDys-2 (\url{https://newton.spacedys.com/astdys/}), and fit using the \cite{BK} orbit fitting code. For each epoch, we also used the online version of Find\_orb (\url{https://www.projectpluto.com/fo.htm}) to produce a comparison fit and uncertainty in the same reference frame. Note the wide range in estimated uncertainties and the fact that the best-fit $a$ value can differ by up to 50\% of the 1-$\sigma$ $a$ range when using different fitting procedures. 
    These relatively discrepant uncertainty estimates are typical for distant TNOs.
    \label{t:sl102}}
\end{deluxetable}

\section{Numerical investigation of dynamics near the observed objects}\label{s:sims}

To explore the range of possible dynamical evolution of observed TNOs, previous studies have carried out long timescale (typically 10~Myr) N-body orbit integrations of many clones of each object's orbit generated with initial conditions within its orbit-fit uncertainties (e.g., \citealt{Gladman:2008}). 
Here we take a different approach: we make maps of the dynamical neighborhood of each TNO by constructing Poincar\`e return maps of a bundle of orbits in proximity to a TNO's orbit.
Resonances can be clearly identified in these maps using quite short integrations ($\lesssim1$Myr); we have extended our integrations to typical timescales of 5-10~Myr to identify relatively stable libration zones (discussed further below).
As we will show, a relatively sparse sampling of initial orbits is required to derive a useful picture of the nearby dynamical landscape.
We take advantage of the fact that, for most TNOs, some orbital parameters (such as the inclination, $i$, and longitude of ascending node, $\Omega$) are much better constrained than others.
In our analysis, we adopt the best-fit values of $i$, $\Omega$, and perihelion distance $q$, and then explore the dynamical landscape in the TNO's orbital neighborhood by varying the semimajor axis, $a$, and argument of perihelion, $\omega$; we note that $\omega$ is also typically well-constrained for observed TNOs, but scanning over both $a$ and $\omega$ in our simulations is key to probing nearby resonant structures.
Consequently, with our particular choice of variables for the 2D projections of the Poincar\`e maps (described in the next section), it becomes possible to very directly assess the role of Neptune's MMRs on each observed TNO.
Moreover, the proximate dynamical landscape is fixed and unvarying, even as future observations refine and update a TNO's orbit. 
With a Poincar\`e map of a TNO's neighborhood in hand, it becomes possible to identify quickly the need (or not) to re-classify that TNO if its orbit is updated with new observations.

Poincar\'e return maps have previously been used primarily with simple models, such as the circular planar restricted three body problem, where the return map is a two dimensional surface and thus can be conveniently visualized in 2D plots \citep[e.g.][]{Henon:1966,Winter:1997}. 
These plots readily delineate the zones of stable resonant librations, stable non-resonant zones, as well as chaotic zones. 
For three degrees of freedom and multiple perturbers (on non-circular and non-coplanar orbits), Poincar\'e return maps have not previously been thought to be of much use because there is an insufficient number of conserved quantities to allow us to visualize the dynamical landscape in two dimensional surfaces. 
However, in this work we demonstrate their usefulness when they are constructed with well-chosen criteria and the high-dimensional return map is rendered in 2D projections of well-chosen pairs of orbital parameters. 
This allows us to directly assess whether a distant TNO is likely influenced significantly by Neptune's MMRs or if it can be reliably classified as `detached'.

\subsection{Methods}\label{ss:methods}

Our approach is based on a recently developed implementation of Poincar\'e return maps of the circular planar restricted three body model \citep{Wang:2017,Lan:2019,Malhotra:2021}.
We first briefly describe this method for the case of the simple model and then describe the additional modifications we have made to extend it to the case of the three dimensional restricted N-body model, that is a massless test particle moving under the perturbations of N-1 point-mass bodies under mutual Newtonian gravitational forces.

\begin{figure*}[ht]
    \centering
    \begin{tabular}{c c}
    \includegraphics[width=3.25in]{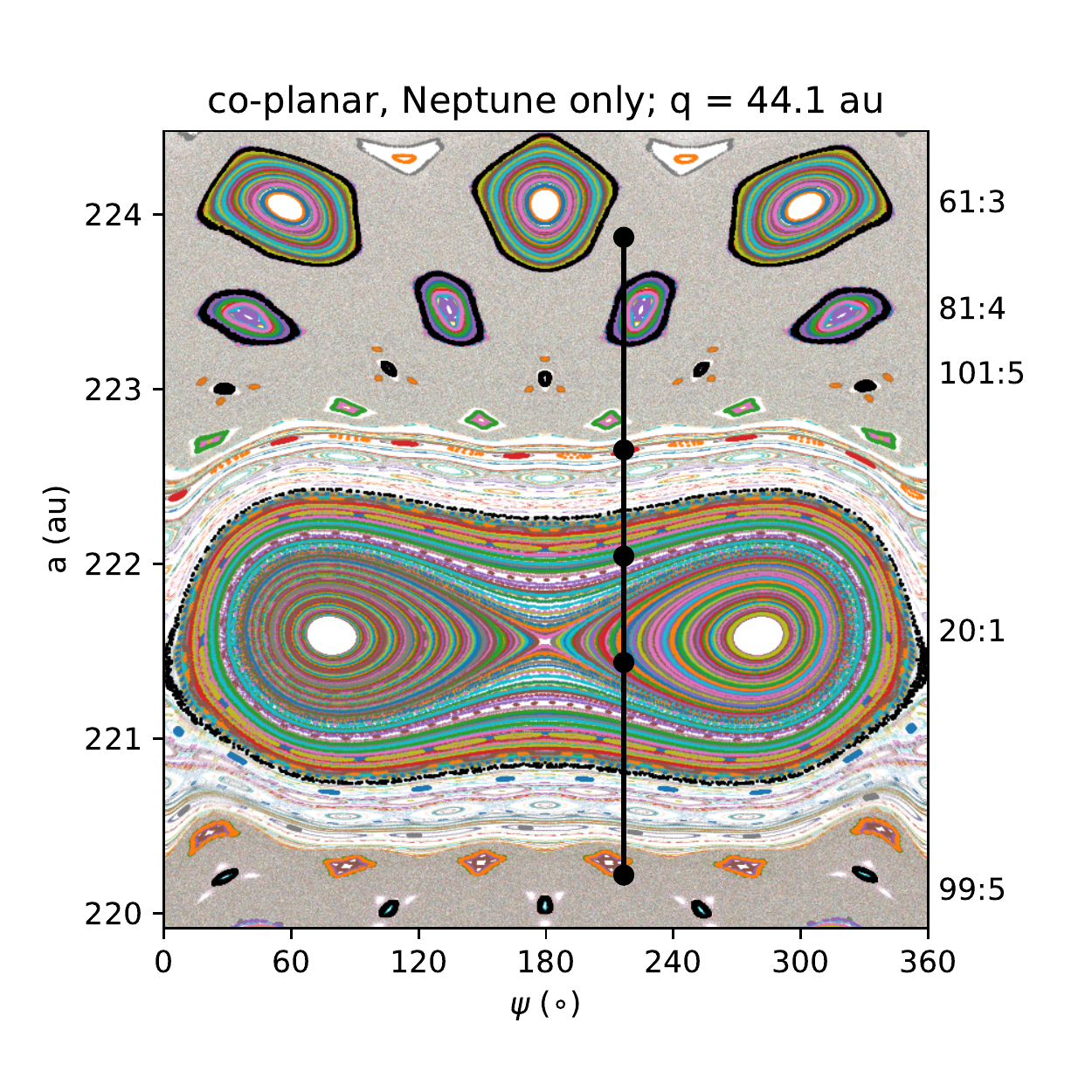} & 
    \includegraphics[width=3.25in]{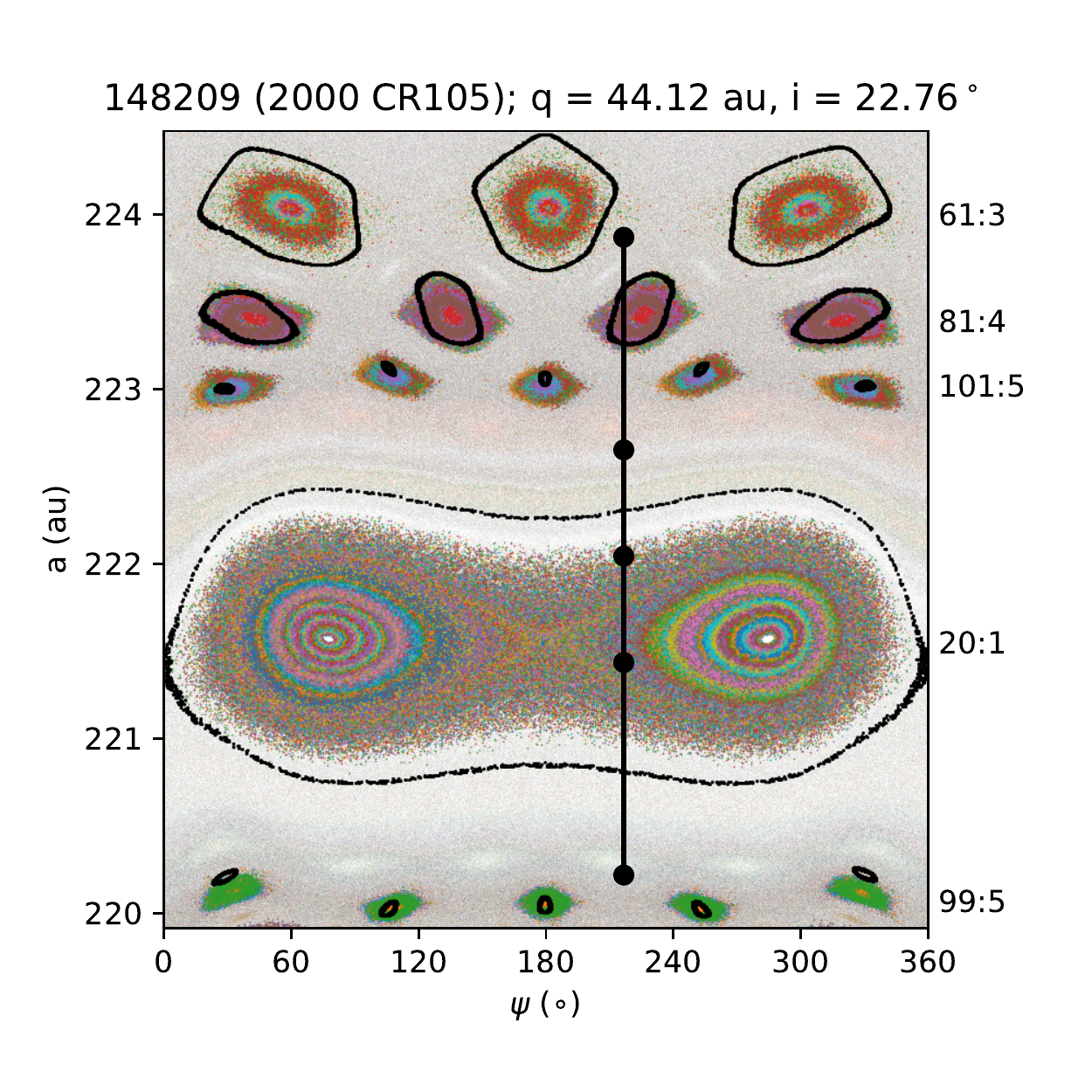} \\
    \end{tabular}
    \caption{Semimajor axis vs $\psi$ Poincar\'e maps in the circular planar restricted three body problem (left panel) and in simulations with all 4 giant planets on their present orbits and test particles on inclined orbits (right panel) for a region of phase space near the orbit of TNO 148209 (2000 CR105). 
    The large black dots connected by black vertical lines show then TNO's best-fit orbit (center circle) along with 1-$\sigma$ and 3-$\sigma$ uncertainties in $a$; the uncertainty in $\psi$ is negligible). 
    The plots show the semimajor axis vs $\psi$ for test particles (different particles shown in different colors) when they pass through perihelion. 
    To best highlight the resonant structures, particles that remain in resonance with Neptune for more than 1000 resonant cycles are plotted with the largest, most opaque points; particles librating for 750--1000 resonant cycles are plotted in slightly smaller, slightly more transparent points, as these particles' paths temporarily stick near the resonant islands before dispersing into the non-resonant phase space.
    Other, generally non-resonant, chaotic  particles are shown in small, more transparent points. 
    The resonant boundaries obtained from the three body problem (the librating resonant test particles with the largest $a$ and $\psi$ variations) are shown as black curves in both panels for the labeled mean motion resonances with Neptune; the simplified problem captures the widths of the strongest resonances quite well.}
    \label{f:ex}
\end{figure*}

In their implementation of Poincar\'e return maps in the circular planar restricted three body model, \cite{Wang:2017} and \cite{Lan:2019} carried out numerical orbit integrations for many initial conditions (with the same Jacobi integral) and recorded the state vector of the test particles at every pericenter passage.
They then plotted 2D maps in two pairs of orbital parameters, $(\psi,a)$ and $(e\cos\psi,e\sin\psi)$, where $a,e$ are the semimajor axis and eccentricity, and $\psi$ is the difference in true longitude between the test particle and the planet.  
In this simplified model, $\psi$ is the physical angle between the barycentric position vectors of the planet and particle when the latter passes perihelion. 
(For a $p:q$ exterior resonance, $\psi$ is related to the usual critical resonant angle, $\phi=p\lambda - q\lambda_N - (p-q)\varpi$, by noting that $\lambda=\varpi$ when the particle is at perihelion, so $\phi=q(\lambda-\lambda_N)=q\psi$.)
In these 2D maps, particles at exact resonance will appear sequentially at a discrete set of points (the resonance centers) in the $(\psi,a)$ plane.
Particles librating stably with a non-zero libration amplitude will trace out smooth, closed (bounded) curves around each resonance center; collectively these will appear as a chain of resonant islands. 
Those particles not in resonance but in regular, quasi-periodic motion will trace out smooth 1D curves in which $\psi$ circulates through all values 0--360$^\circ$. 
Those in chaotic motion will random-walk over a 2D region in the parameter plane. 
An example of such a Poincar\'e map in the $(\psi,a)$ plane is shown in 
the left panel of Figure~\ref{f:ex} for orbits near Neptune's 20:1 exterior mean motion resonance in the circular planar restricted three body model.
Traces of individual particle initial conditions are plotted using different colored dots; the resonant paths are emphasized using slightly larger, more darkly colored dots.
The large, two-island structure at 221.6~au indicates the libration zone of the 20:1 resonance. 
Other chains of islands (with two, three, four, or more islands) are also visible in this neighborhood.

The N:1 exterior resonance exhibits a more complex dynamical structure than other exterior resonances, so it is worth explaining in some detail.
Particles librating in an N:1 exterior resonance have a single perihelion passage in a synodic period and can exhibit librations about three different centers, forming three different libration zones: two zones of so-called asymmetric librations, and a zone of symmetric librations. 
In the example shown in Figure~\ref{f:ex}, the asymmetric resonant particles come to perihelion with values of $\psi$ that librate around either $\approx70^\circ$ or $\approx290^\circ$; these are the two tear-drop shaped asymmetric islands in the left panel of Figure~\ref{f:ex}, and they are traced by independent sets of test particles. 
The symmetric resonant particles come to perihelion with $\psi$ librating around $\psi=180^\circ$, tracing a path of large libration amplitude that encloses both asymmetric islands.

Neptune's other exterior resonances are simpler in their dynamical structure.
For an $N:k$ exterior resonance, resonant particles sequentially visit each of a chain of $k$ resonant islands over their $k$ perihelion passages each resonant cycle. 
In Figure~\ref{f:ex}, the second strongest resonance (after the 20:1) is the 61:3 near the top of the left panel.
Particles in this resonance have three perihelion passages per resonant cycle;  in the Poincar\`e return map, they trace curves which librate around centers at $\psi=60, 180, 300^\circ$, creating the three-island chain near $a=224$~au. 
Particles in the 81:4 resonance trace out the four-island chain just below the three-island chain of the 61:3 resonance.
We observe a `sea' of points in-between the resonant islands; these are generated by initial conditions of chaotic orbits. 
Maps from this simplified, coplanar restricted three body model of the Sun--Neptune--test particle system, like those in the left panel of Figure~\ref{f:ex}, can help illustrate the potential for resonant interactions near an observed TNO's orbit.
However, to get the most accurate picture, we must include the effects of the other giant planets, relaxing the assumption of circular planet orbits and of co-planarity.

With a full N-body model, we adopt a procedure inspired by but modified from that of the simple three body model described above.
We construct Poincar\`e return maps for bundles of test particles evolving under the influence of the Sun and all four giant planets by recording the state vectors at each perihelion passage. 
Rather than initializing all particles in the map with the same Jacobi constant as in the simplified model, we instead initialize them with fixed perihelion distances (this difference is discussed further in Appendix~\ref{appendix}).
We then make 2D plots of the barycentric semimajor axis versus a modified version of the relative longitudes, $\psi$, defining 
$\psi=\lambda-\lambda_N$, where $\lambda =M+\omega+\Omega$ (the sum of mean anomaly, the argument of perihelion, and the longitude of ascending node) is the mean longitude of the test particle, and $\lambda_N$ is the mean longitude of Neptune. 
The right panel of Figure~\ref{f:ex} shows such a Poincar\`e map near Neptune's 20:1 resonance with the full N-body model and test particles on significantly inclined orbits; for comparison, we overlaid in black the maximum extent of the libration zones of the resonant islands found in the simple model. 
We observe that the Poincar\`e map of the N-body model is a `blurred' version of the map seen in the simple model, but the dominant resonant structures are still evident. 
To more clearly visualize the resonant islands and separate them from the sea of non-resonant orbits in Figure~\ref{f:ex} and our later plots, we use different point sizes and transparencies depending on dynamical behavior. 
We show the paths of non-resonant particles using the smallest, most transparent points. 
Particles in stable libration zones, which we define as particles that complete at least 1000 resonant cycles in our simulations, are shown with the largest, most opaque points.
Quasi-resonant particles, which we define as completing between 750 and 1000 resonant cycles before dispersing into the chaotic sea, are shown as medium sized, slightly transparent points.
These distinctions are by necessity somewhat arbitrary, but we have found them to provide more informative visualizations of the phase space.
The resonance structures identified in our N-body Poincar\`e maps show much similarity with but also some interesting differences compared 
with those in the simplified model.
It is evident that the phase space near resonances is affected by both inclination and by the perturbations of the additional giant planets; a detailed analysis of how different resonances behave as a function of inclination and under perturbations from the other planets is deferred to future studies.
See Appendix~\ref{appendix} for additional validation of our modified Poincar\`e mapping approach.

The vertical line in both panels of Figure~\ref{f:ex} indicates the best-fit and 3-$\sigma$ uncertainty range of the semimajor axis for the observed TNO 148209 (2000 CR105) plotted at its observed $\psi$ value. 
(Note that for all the observed TNOs considered here, $\psi$ is very well determined, with its uncertainty being less than the thickness of the line; we thus ignore the uncertainty in this parameter.) 
With this comparison, we can conclude that 148209's best-fit orbit is consistent with libration in the 20:1 resonance, although the orbit-fit uncertainties must be reduced in order to have high confidence that this is a resonant TNO.
Below we describe in detail how we construct these maps for each of the objects listed in Table~\ref{t:objs}, including the choice of initial conditions for the bundles of proximate orbits near an individual TNO.

The vast majority of the observed objects in Table~\ref{t:objs} are currently near their closest approach to the Sun. 
Their perihelion distances are thus observationally well constrained, so we will take them to be fixed in our simulations.
We similarly take their inclinations and longitudes of ascending node to be fixed at their best-fit values because these orbital elements converge very quickly and have small observational errors. 
The objects' arguments perihelion are similarly well-constrained, but we take this as a free parameter in our simulations in order to map the full range of potential resonant interactions near the observed orbit.
The semimajor axes of the objects contain the bulk of the orbital uncertainties, so we consider initial semimajor axes spanning their full 3-$\sigma$ ranges (except in a few instances where these ranges are very large, in which case we restrict ourselves to the 1-$\sigma$ range). 

For each of the objects in Table~\ref{t:objs}, we initialize test particles at perihelion (mean anomaly $M=0$) at initial barycentric semimajor axes corresponding to every N:$\{1,2,3,4,5\}$ resonance with Neptune within the observed uncertainty range and slightly beyond on either side.
We limit ourselves to this set of resonances because for the large eccentricities in the high-$a$ TNO population, N:1 resonances are expected to be the strongest local resonances, followed by N:2 then N:3 resonances, and so on (see, e.g., \citealt{Gallardo:2006}'s estimates of resonance strengths in the scattering TNO population and discussion of high-$e$ resonance order in \citealt{Pan:2004}).
For 2015 RX245 and 2016 SD106, we only initialize particles in the N:1 and N:2 resonances due to the large semimajor axis uncertainties to be covered and the fact that the resulting maps show only weak stability in the N:2 resonances. 
Similarly, for 2010 GB174, we limit ourselves to N:1, N:2, and N:3 resonances. 
We also limit the study of the regions near 2010 GB174  and 2018 AD39 to only their 1-$\sigma$ $a$ ranges due to their large orbital uncertainties and the large number of resonances with stable zones in their regions (both of which increase the computational cost of constructing the maps).
We similarly limit the $a$ ranges of the maps near 2012 VP113 and 2015 KG163 to just the 1-$\sigma$ range because the smaller range is sufficient to tell us resonances are unlikely to be important drivers of their dynamical evolution.

The test particles for each considered resonance for each TNO are given $q,i,\Omega$ values identical to the observed object's best-fit orbit.
We then set $\omega$ to a range of values such that $\psi = \lambda-\lambda_N$ is in the range $0$--$360^\circ/\{1,2,3,4,5\}$, (the divisor in the range set by the resonance being initialized) with test particles spaced every $1^\circ$ in $\lambda-\lambda_N$; this ensures that initial test particles cover the resonant islands we hope to map. 
For each resonance we also include an additional 100 test particles initialized randomly in $a$-$\psi$ the range $a_0\pm1$~au and $\psi=0$--$360^\circ/\{1,2,3,4,5\}$, where $a_0$ is the nominal resonant semimajor axis to account for potential shifts in the center of the resonant islands and to capture the dynamics of nearby higher-order resonances. 
The test particles are then integrated using the \textsc{ias15} adaptive step size integrator in the \textsc{rebound} orbit integration software \citep{rebound,ias15} until they complete 1500 resonant cycles (i.e., 1500 orbits for a particle in an N:1 resonance, 3000 orbits for a particle in an N:2 resonance, etc.) or are scattered more than 5 au outside the initial semimajor axis range of interest.
The largest allowed integration step size in the simulations is set to 0.2 years to ensure we can resolve the time of a particle's perihelion passage with high accuracy (to within $0.05^\circ$) in order to output its state vector at each perihelion passage.
Our simulation length is chosen to maintain computational feasibility while still demonstrating resonant stability over meaningful timescales. 
At our lower semimajor axis limit ($a=150$~au), our simulations span $\{1,2,3,4,5\}$ times $\sim3$~Myr for N:$\{1,2,3,4,5\}$ resonances; the base timescale for N:1 resonances increases to $\sim6$~Myr at $a=250$~au, $\sim10$~Myr at $a=350$~au, and $\sim14$~Myr at $a=450$~au.

\subsection{Results}

Figure~\ref{f:sos-array1} shows the resulting Poincar\`e maps for the objects in Table~\ref{t:objs}.
As with Figure~\ref{f:ex}, we use different point size and transparency to highlight the structure around the resonances.
It is immediately and strikingly evident that stable resonant islands are present in all but one of these maps. 
Only 2015 KG163 is in a phase space that is devoid of even temporarily stable resonant interactions with Neptune. 
With $a\approx680$~au and $q\approx40$~au, this object likely belongs to a class of `diffusing' orbits (see, e.g., \citealt{Bannister:2017}); the test particles in these simulations all experience strong enough changes in energy at perihelion to spread out and fill $a$-$\psi$ space within relatively few orbits.
The phase space surrounding Sedna and 2012 VP113 also stand out in this collection of maps. 
Both these objects have such large perihelia that the non-resonant test particles in the simulations experience almost no drift in semimajor axis over time (as evidenced by the nearly straight traces in $a$-$\psi$). 
While there are resonant islands in the vicinity of both objects, it seems unlikely that either object is strongly perturbed by these MMRs. 
Sedna's observed value of $\psi$ places it firmly outside the 69:1 island in the map, even if the 3-$\sigma$ uncertainty in $a$ sightly overlaps the resonant $a$.
While 2012 VP113's observed $\psi$ is not formally inconsistent with the stable 51:2 resonant islands seen at $a\approx260.5$~au, the resonance islands are very small; the likelihood of it being resonant is therefore small, and the effects of the resonance, even if it is occupied, appear minor.

\begin{figure*}
    \centering
    \begin{tabular}{c c c}
        \includegraphics[width=2.29in]{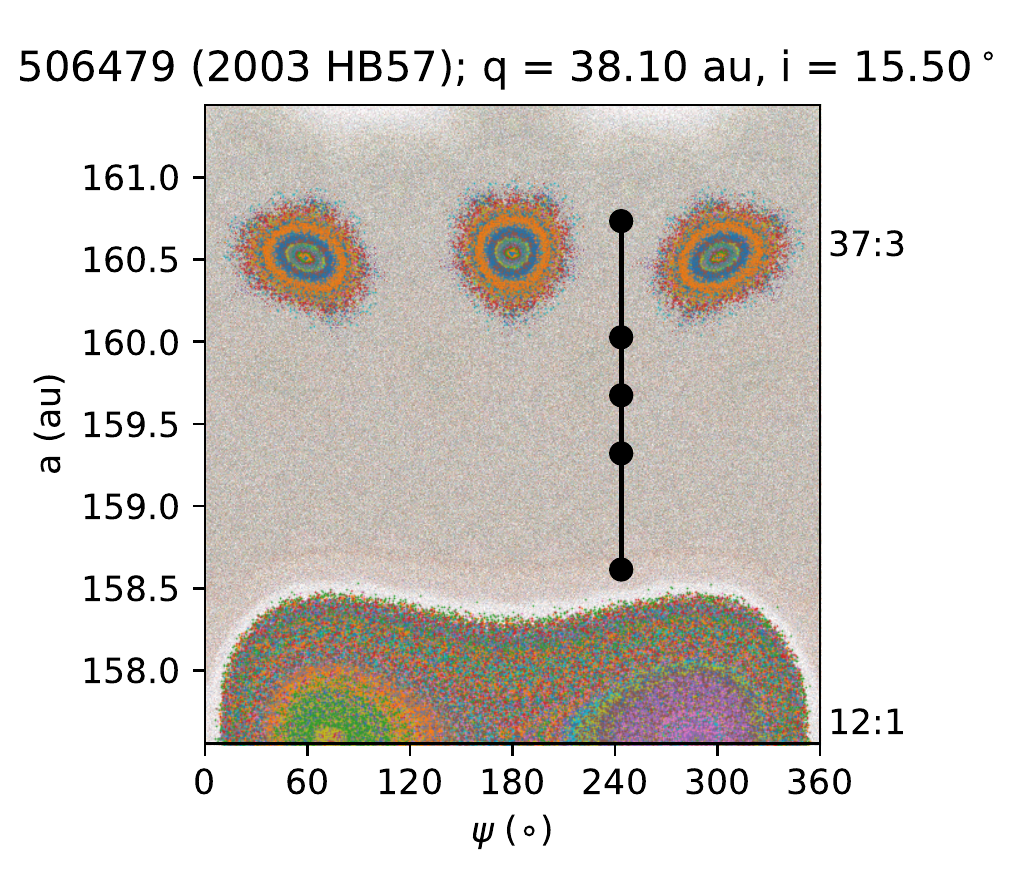} & \hspace{-15pt}
        \includegraphics[width=2.29in]{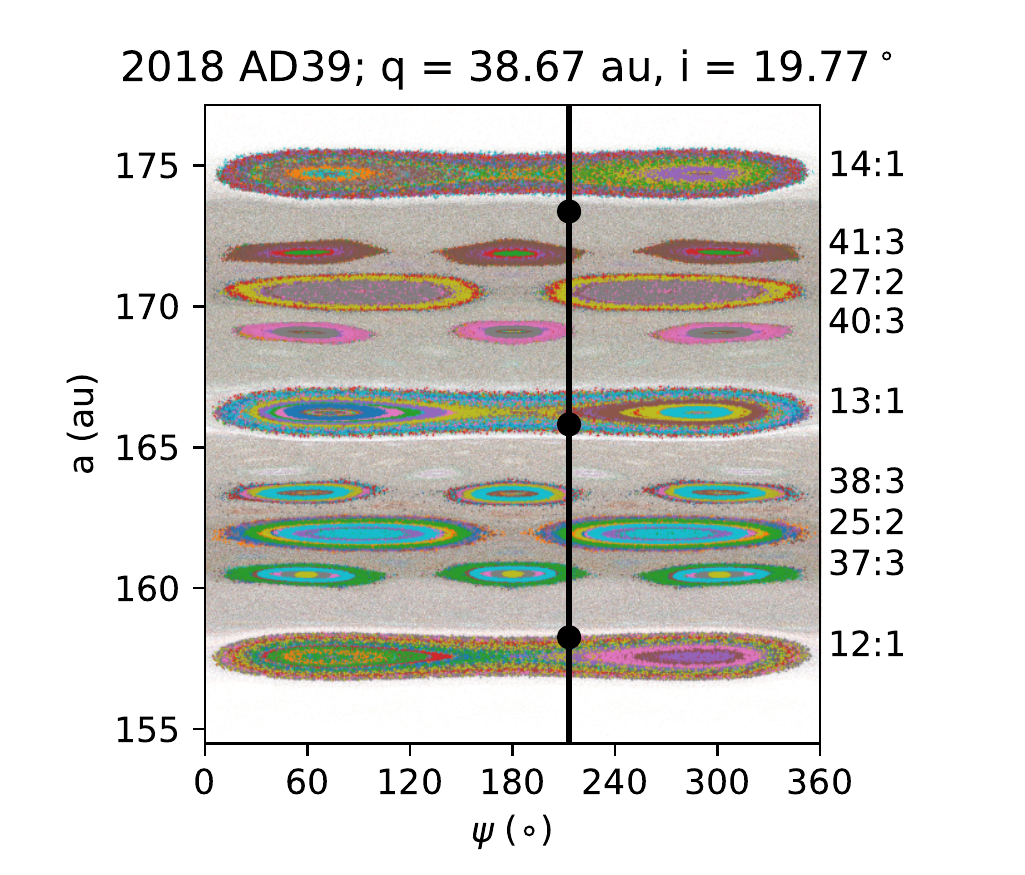} & \hspace{-15pt}
        \includegraphics[width=2.29in]{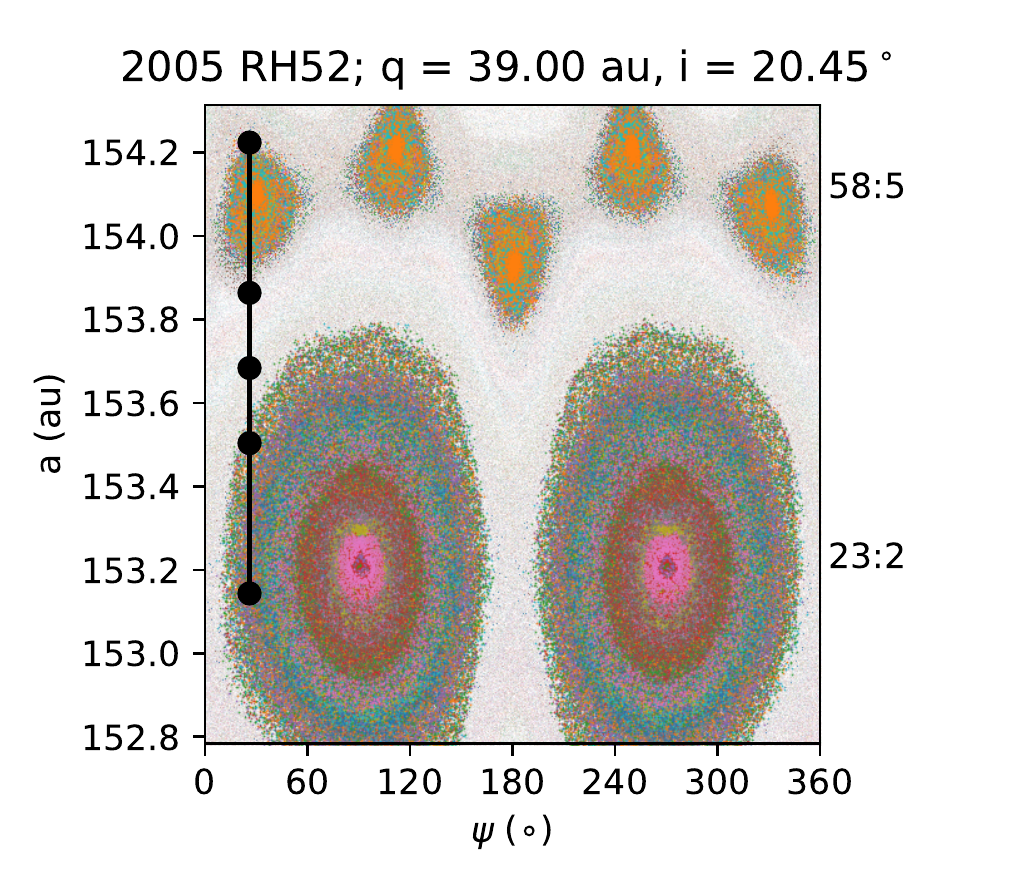} \hspace{-15pt}\\

        \includegraphics[width=2.29in]{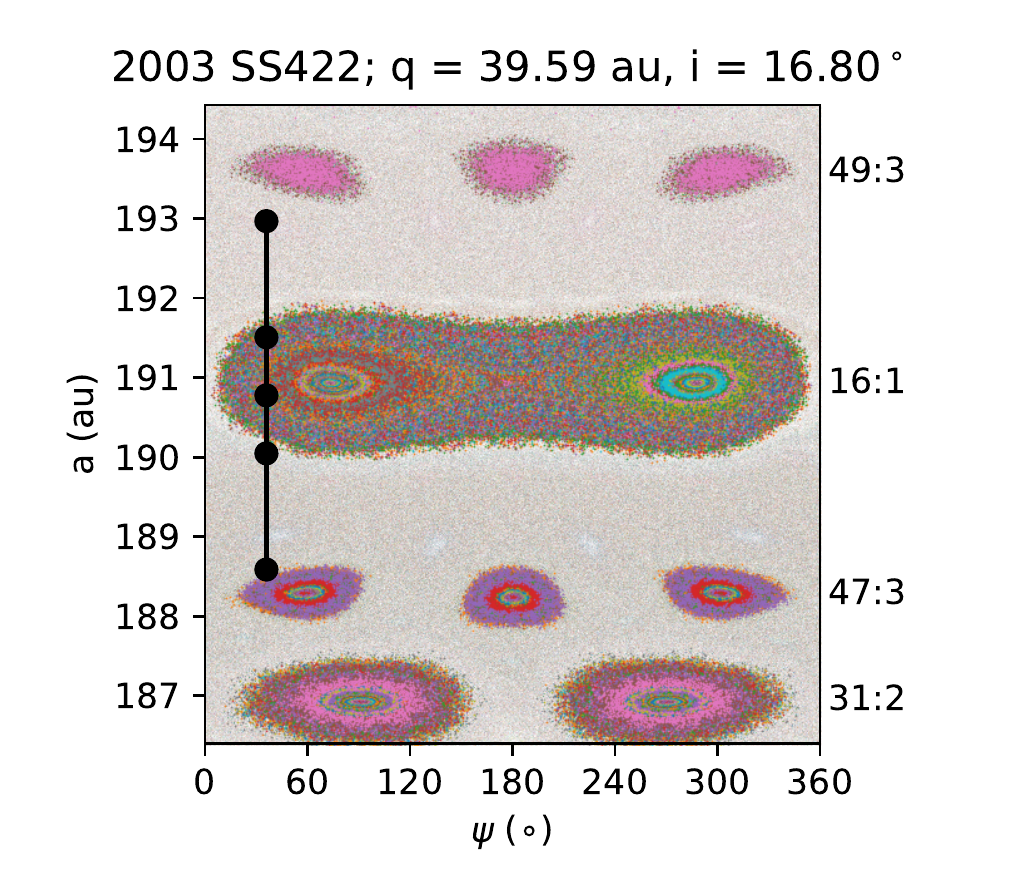} & \hspace{-15pt}
        \includegraphics[width=2.29in]{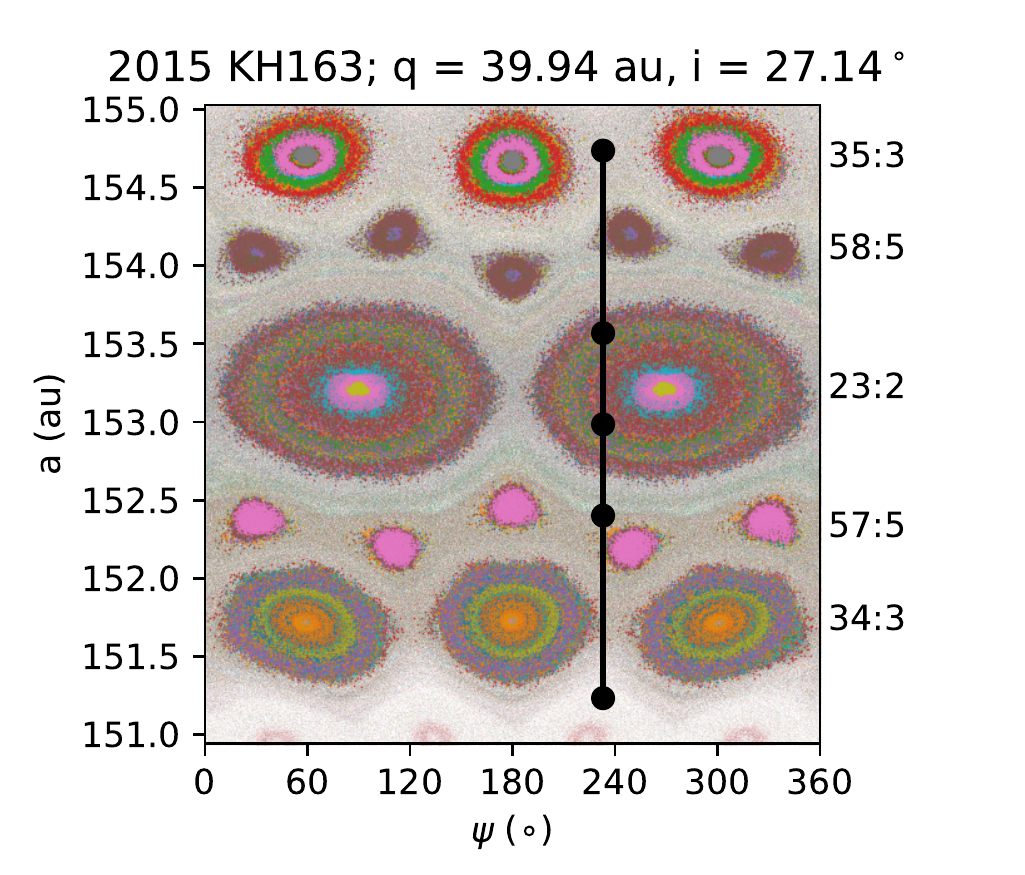} & \hspace{-25pt}
        \includegraphics[width=2.29in]{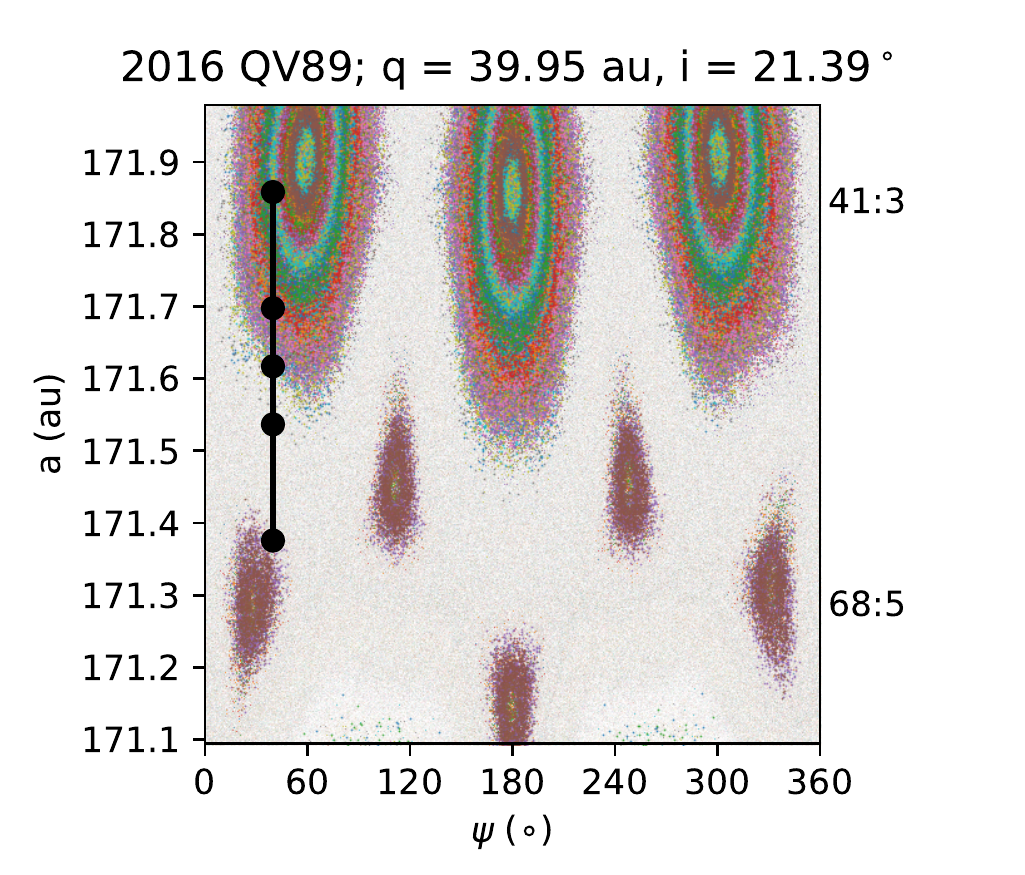} \hspace{-15pt} \\
        
        \includegraphics[width=2.29in]{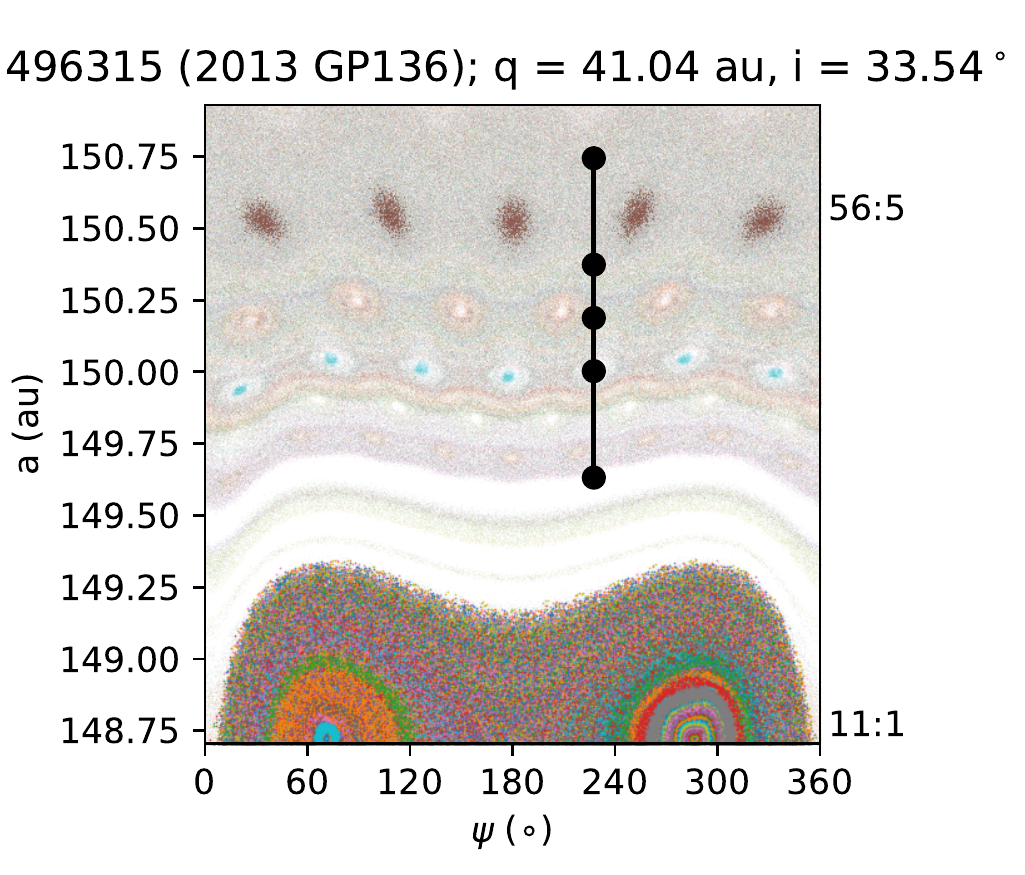} & \hspace{-15pt}
        \includegraphics[width=2.29in]{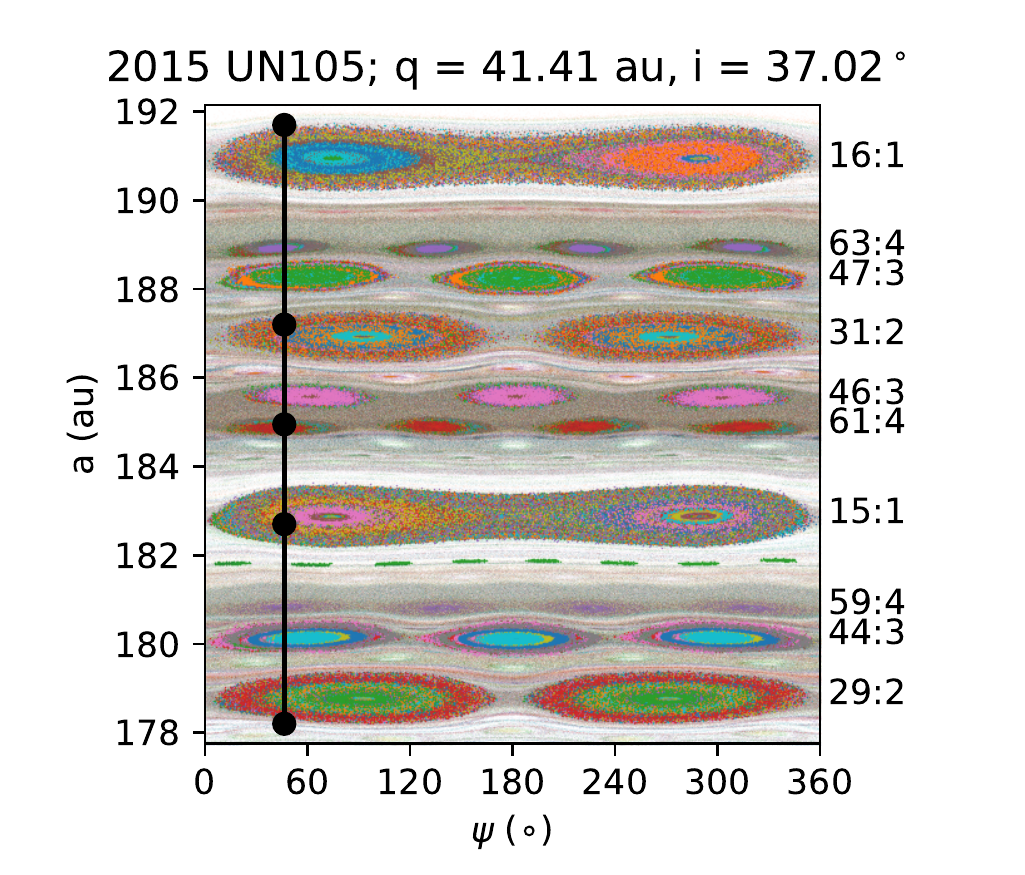} & \hspace{-25pt}
        \includegraphics[width=2.29in]{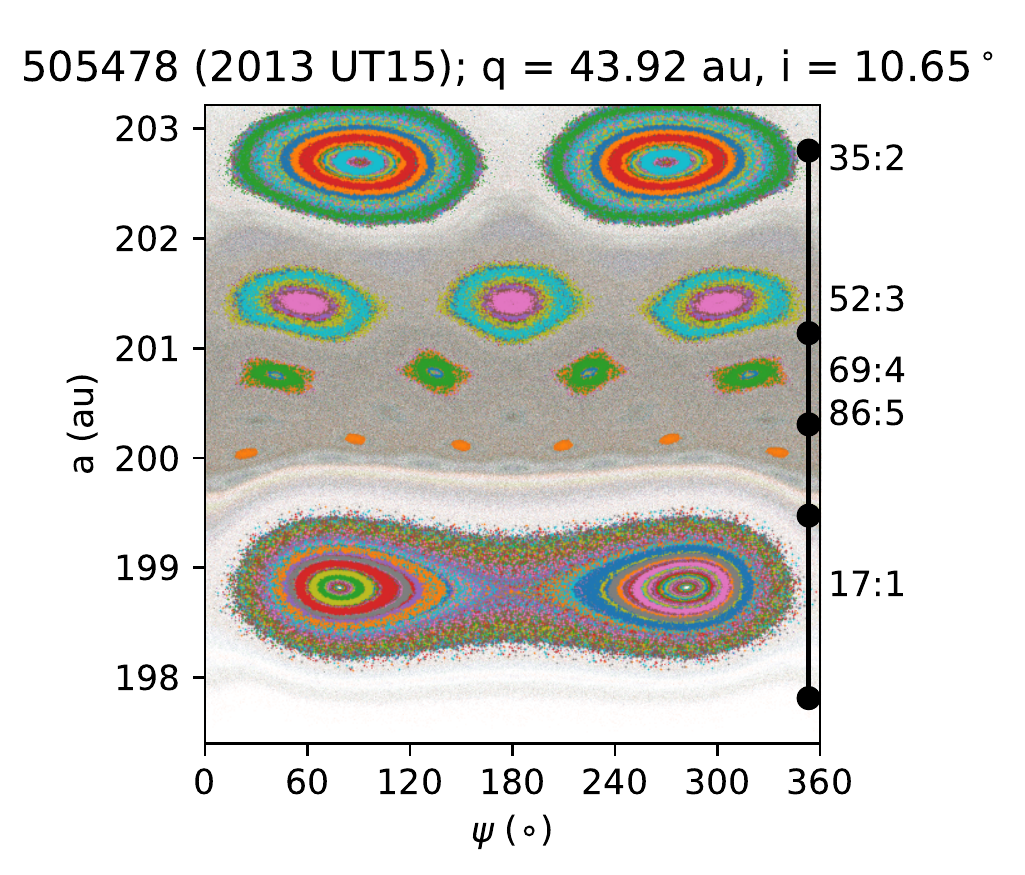}  \hspace{-15pt} \\
        
        \includegraphics[width=2.29in]{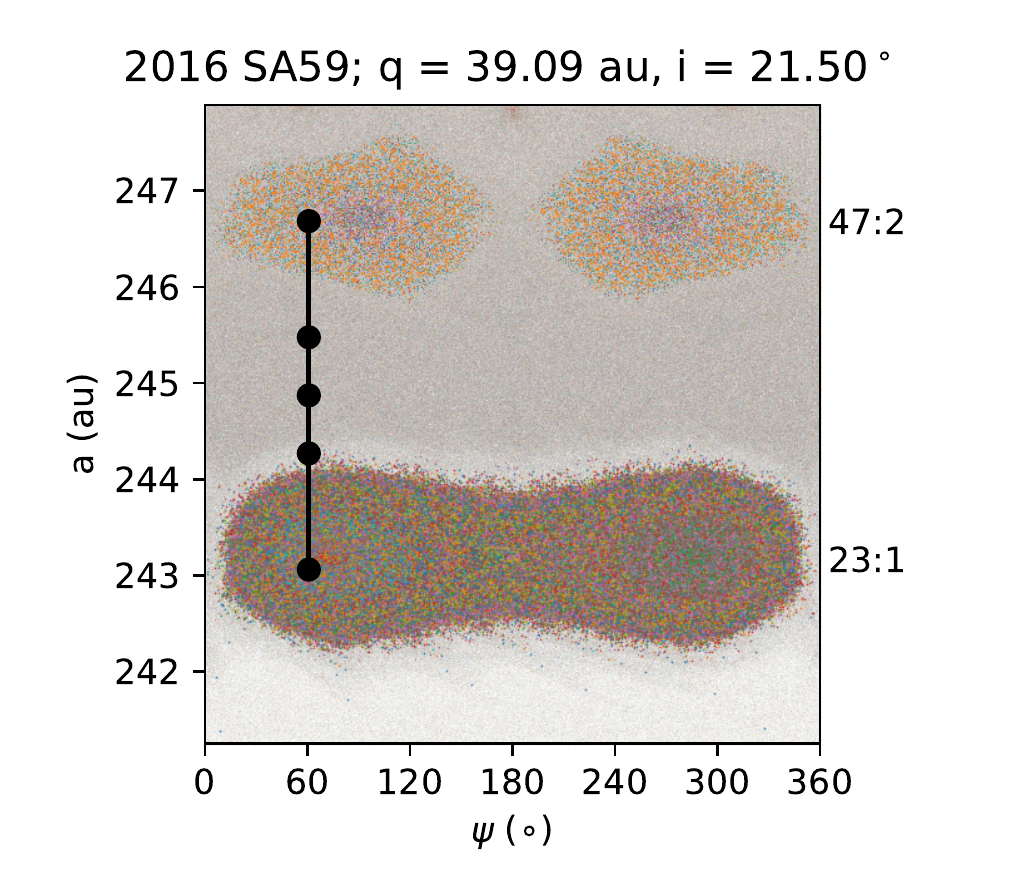} & \hspace{-15pt}
        \includegraphics[width=2.29in]{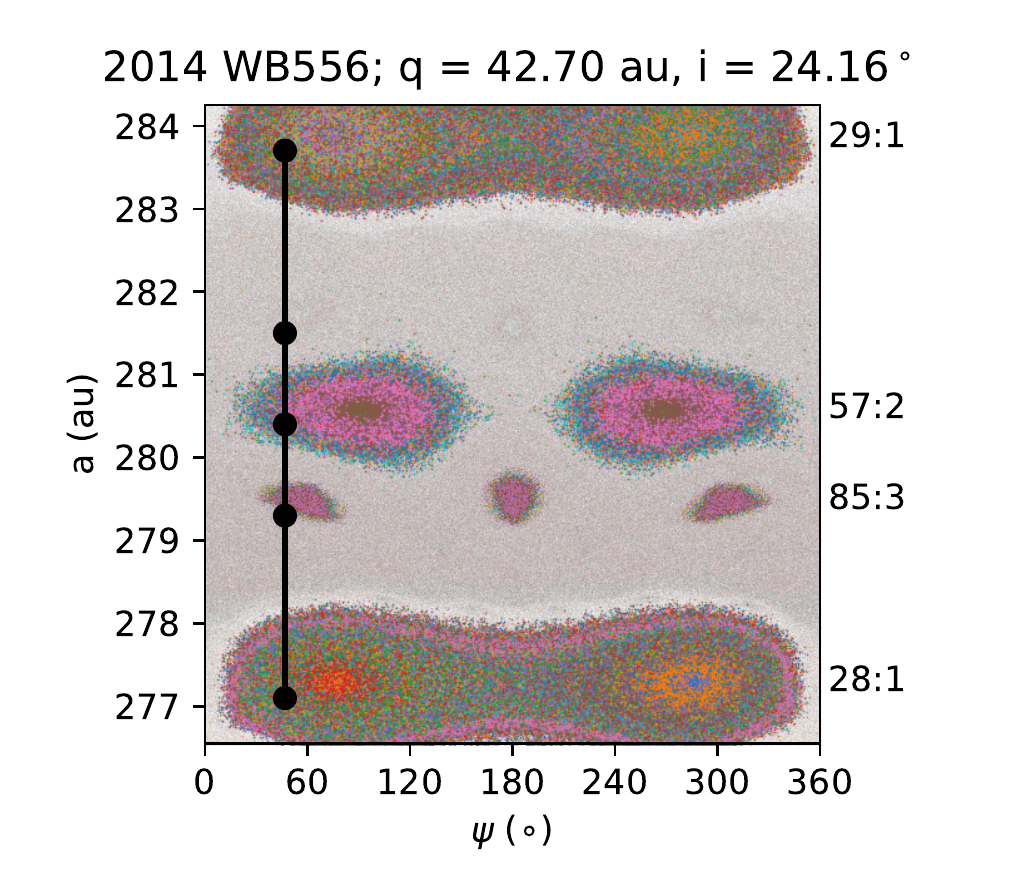} & \hspace{-15pt}
        \includegraphics[width=2.29in]{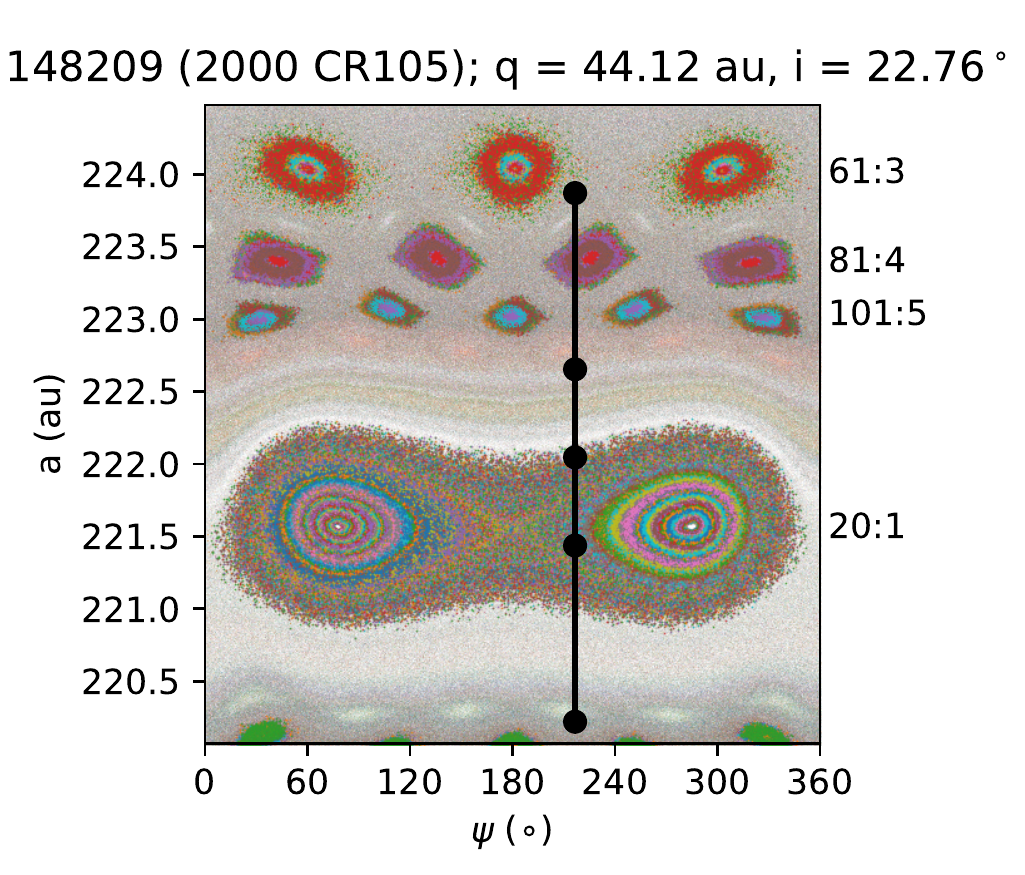}  \hspace{-15pt}\\

    \end{tabular}
    \caption{Poincar\`e maps for test particles on orbits similar to those of the observed high-$a$, high=$q$ TNOs listed in Table~\ref{t:objs}. 
    Each panel shows semimajor axis vs $\psi$ at each perihelion passage for test particles with identical $q$, $i$, and $\Omega$ to the object in the panel title; the test particles are given an expanded range of initial $a$ and $\psi$. 
    Test particles evolve under the influence of all four giant planets, and $\psi$ is defined as $\lambda - \lambda_N$. 
    Each individual test particle is plotted in a different color. 
    Test particles librating in N:1, N:2, N:3, N:4, or N:5 resonances for at least 1000 resonant cycles are shown in the largest, most opaque points; particles that librate for 750-1000 resonant cycles are shown in slightly smaller, slightly less opaque points. 
    Shorter-term resonant particles, higher-order (higher than N:5) resonant particles, and non-resonant particles are plotted with the smallest, most transparent points.
    The black line and circles in each panel shows the observed object's best-fit orbit (middle) and 1- and 3-$\sigma$ uncertainties (taken from JPL horizons); in cases where the uncertainties in $a$ are large and/or span a large number of resonances, only the 1-$\sigma$ range is shown (uncertainties in $\psi$ for the observed objects are very small). }
    \label{f:sos-array1}
\end{figure*}
\begin{figure*}
\renewcommand{\thefigure}{2}
    \centering
    \begin{tabular}{c c c}
        \includegraphics[width=2.29in]{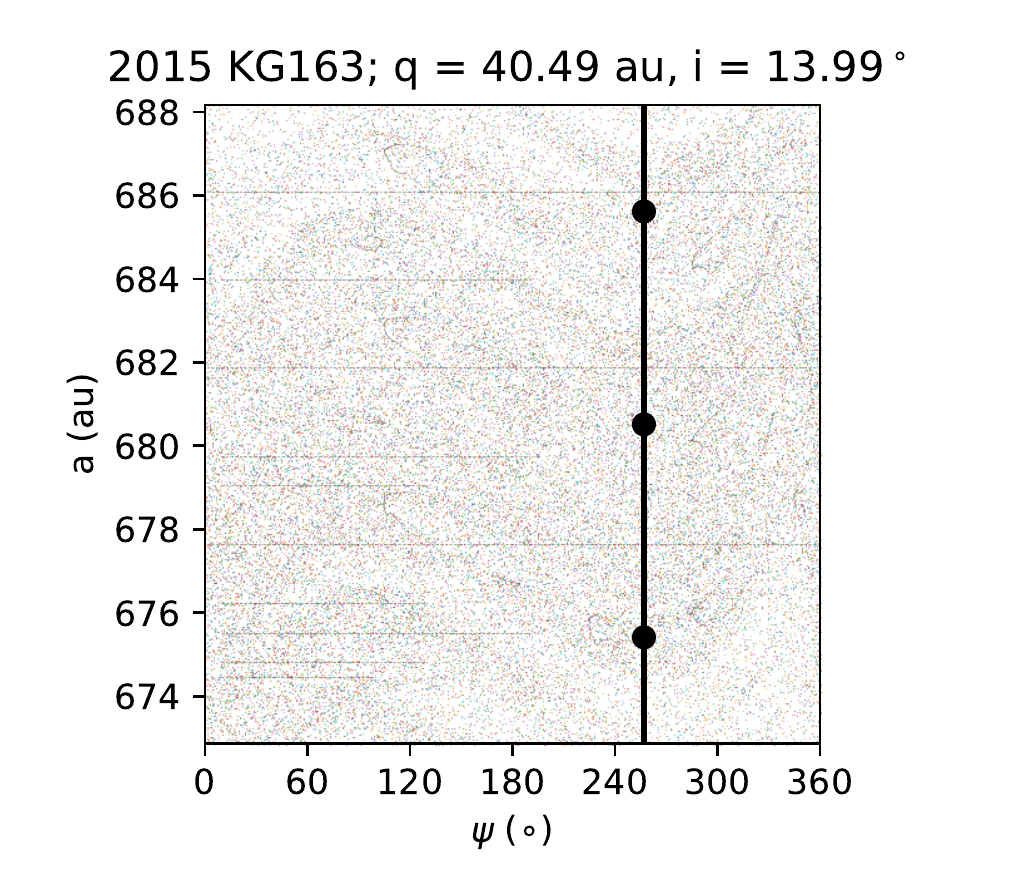} & \hspace{-15pt}
        \includegraphics[width=2.29in]{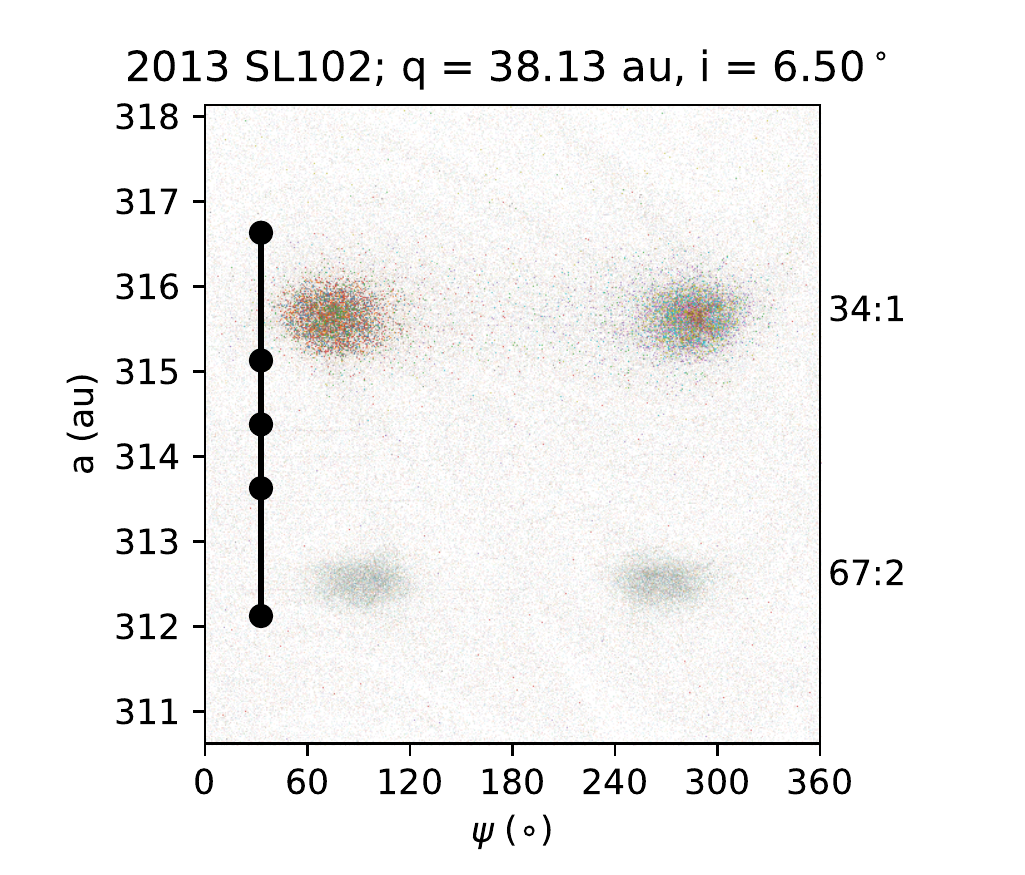} & \hspace{-15pt}
        \includegraphics[width=2.29in]{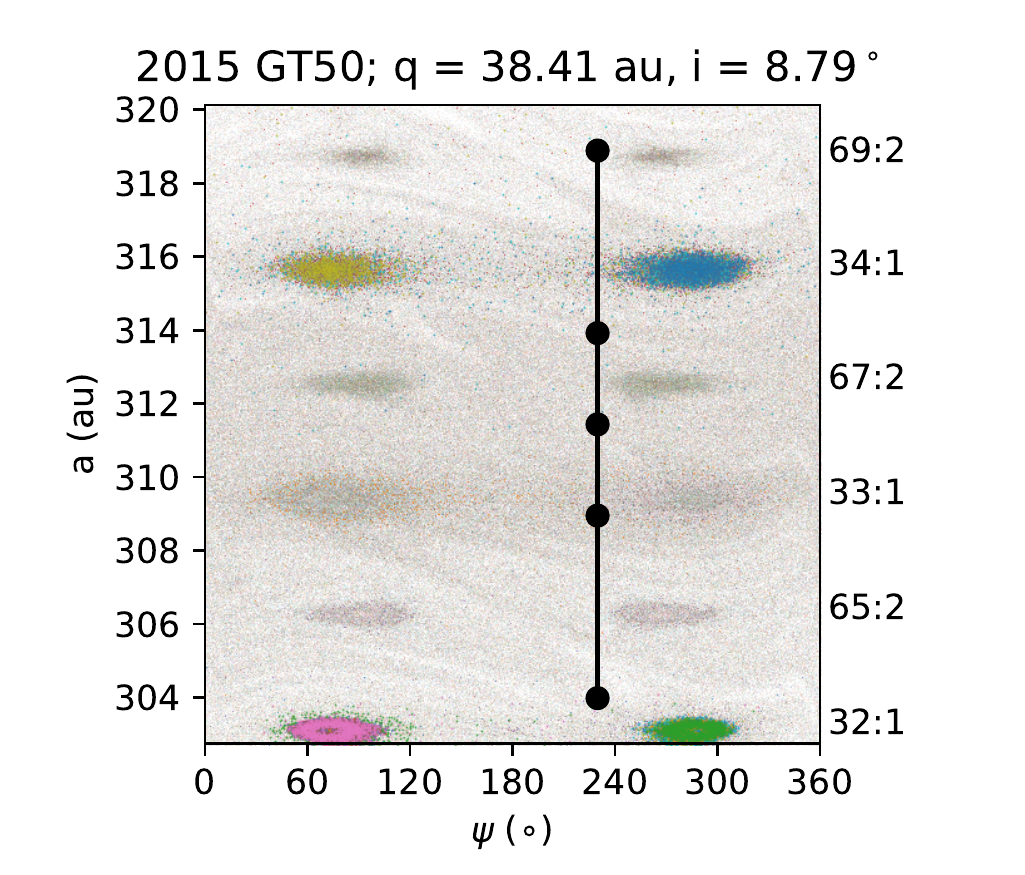}  \hspace{-15pt} \\

        \includegraphics[width=2.29in]{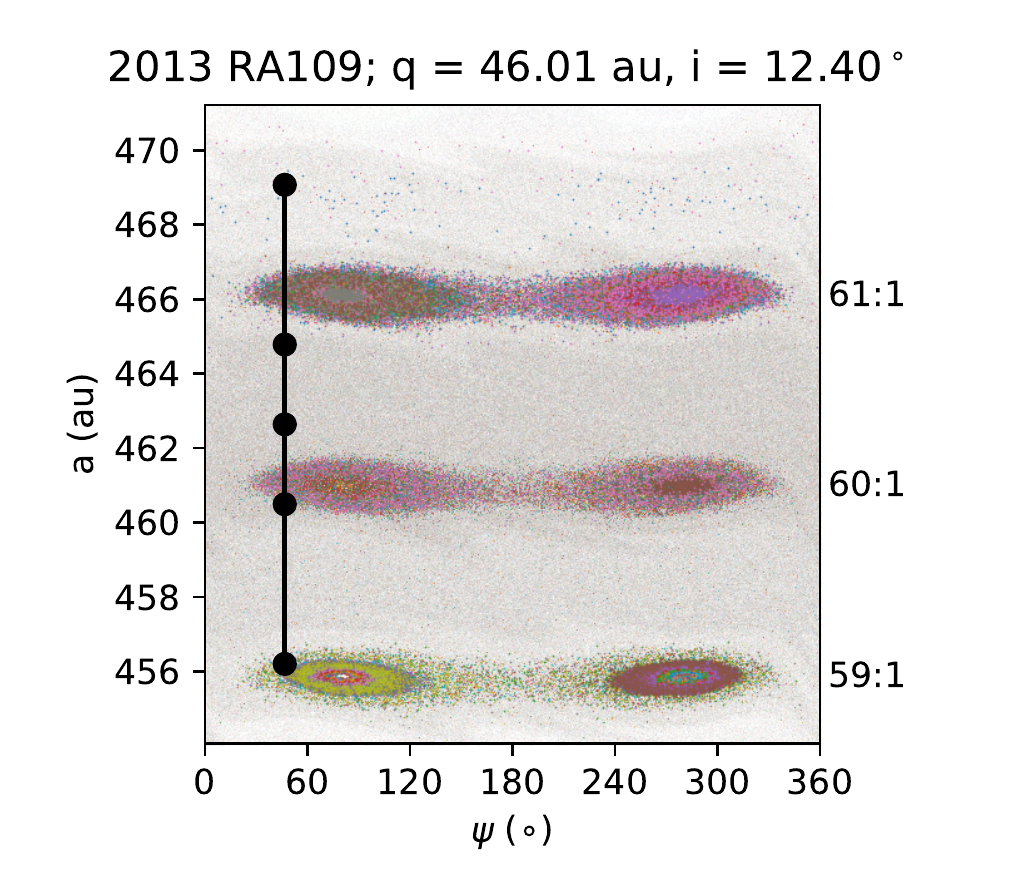} & \hspace{-15pt}
        \includegraphics[width=2.29in]{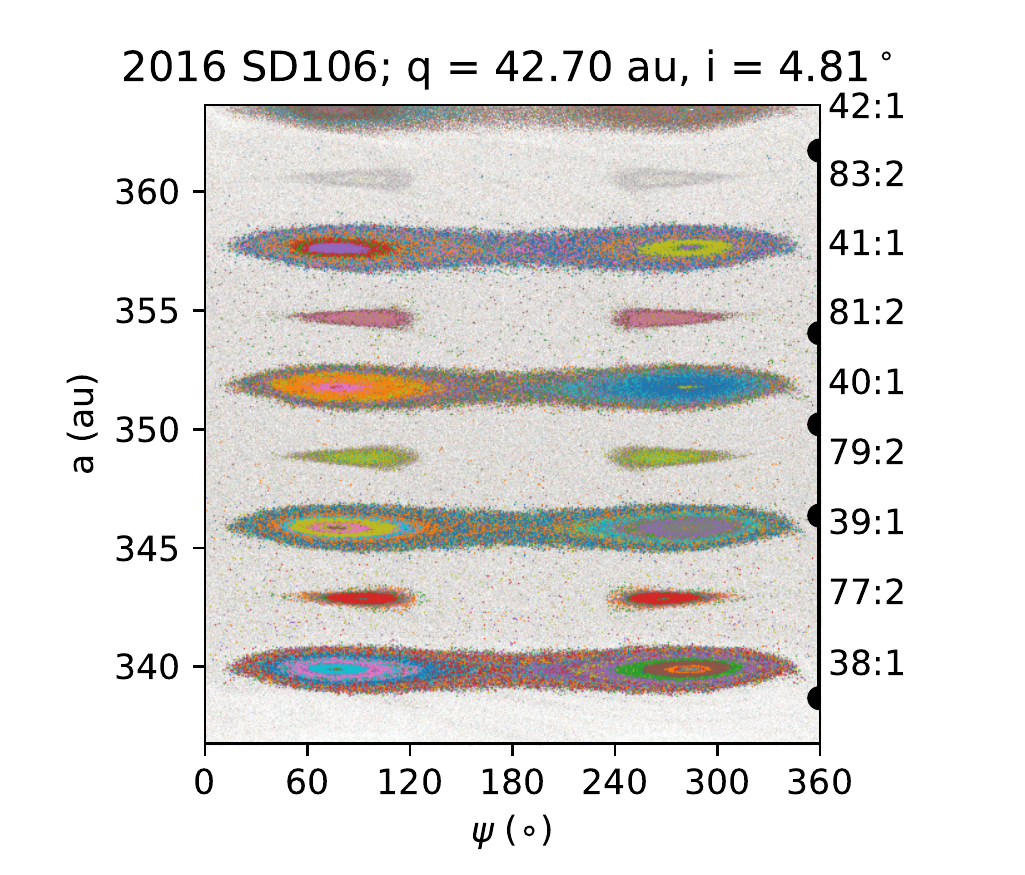} & \hspace{-15pt}
        \includegraphics[width=2.29in]{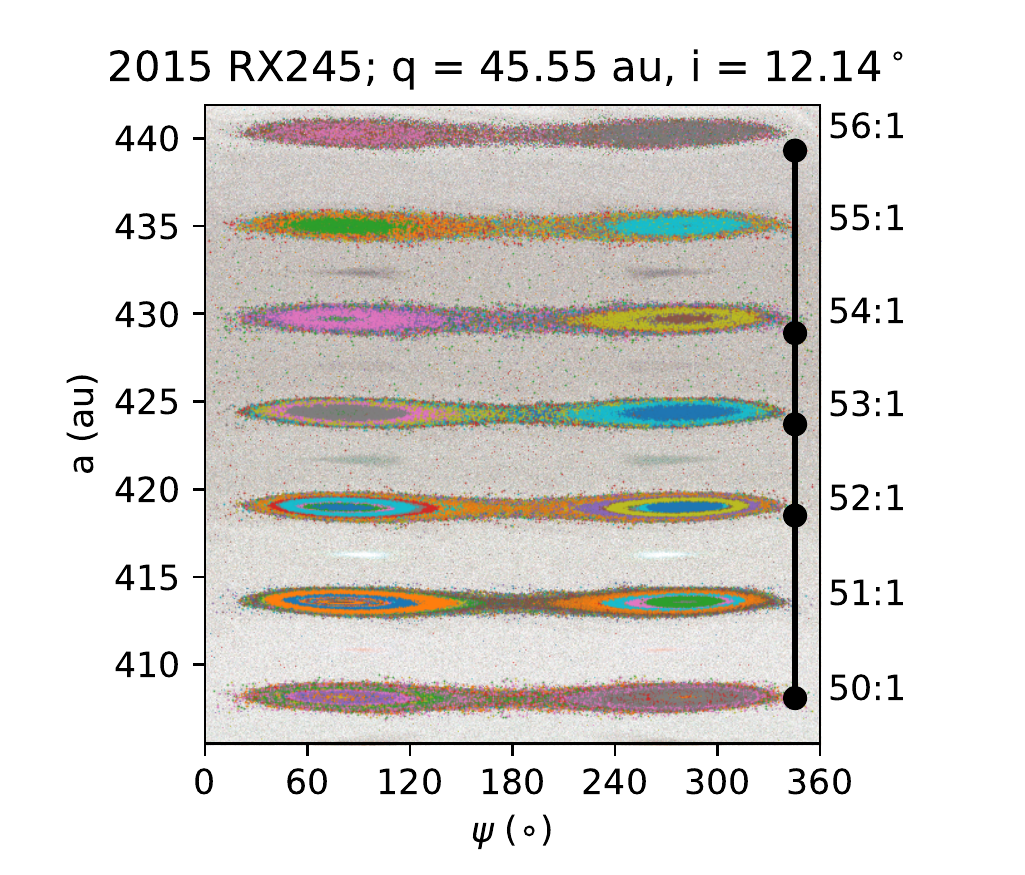}  \hspace{-15pt} \\
        
        \includegraphics[width=2.29in]{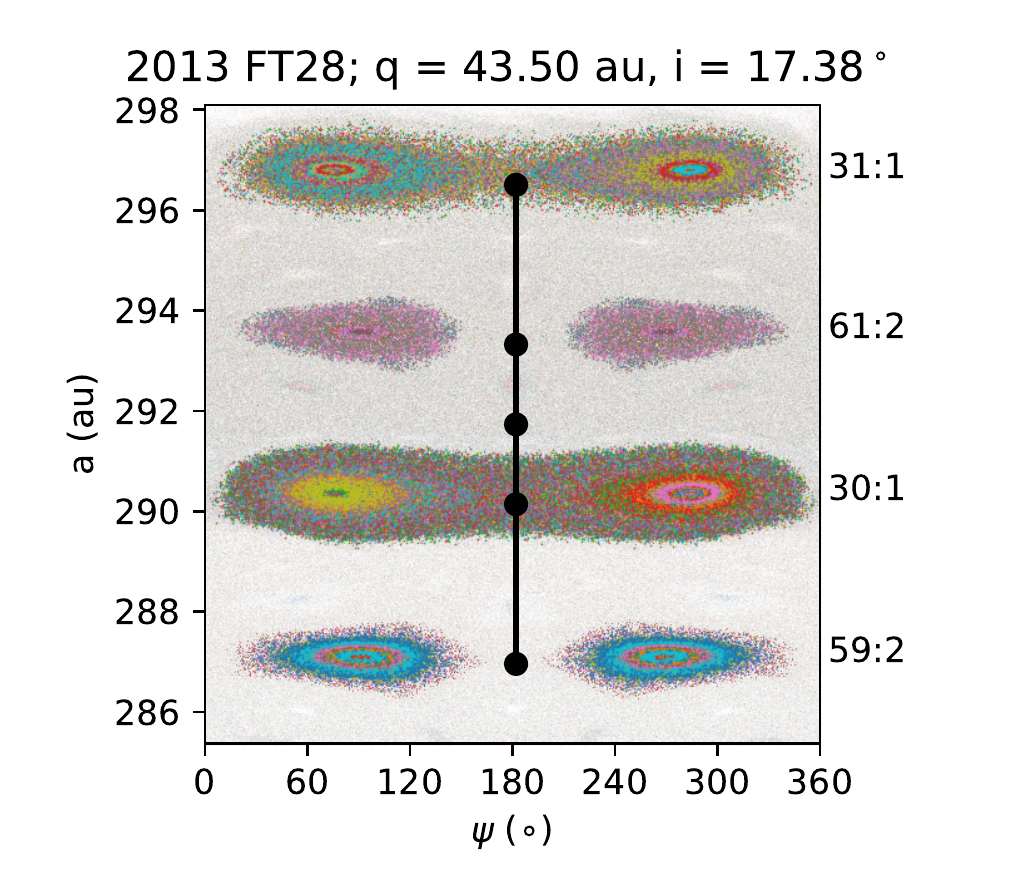} & \hspace{-15pt}
        \includegraphics[width=2.29in]{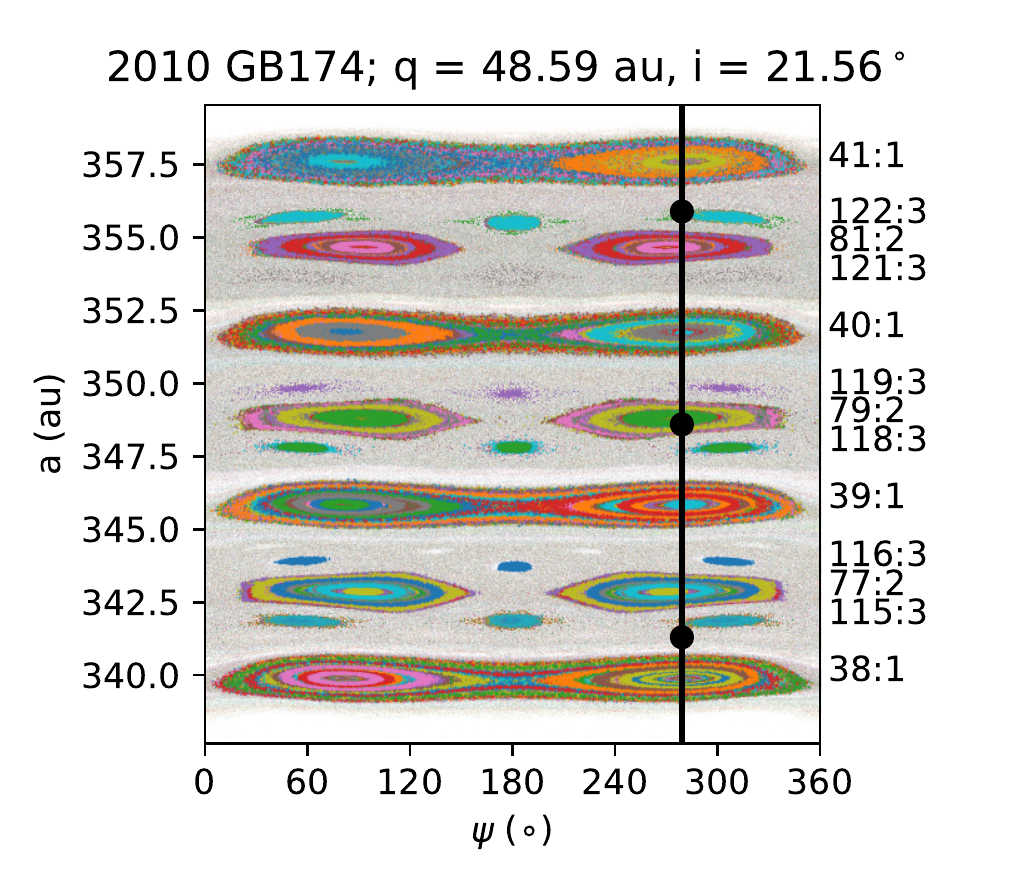} &  \hspace{-15pt}
        \includegraphics[width=2.29in]{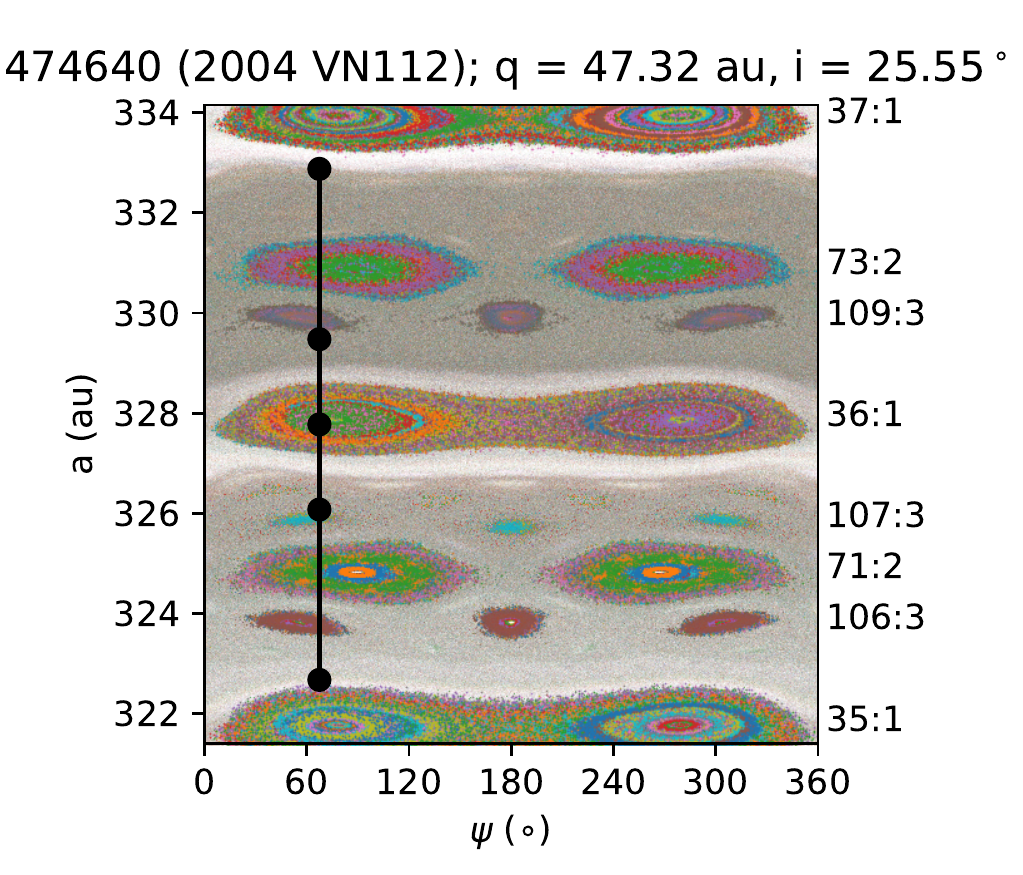}  \hspace{-15pt} \\
        
        \includegraphics[width=2.29in]{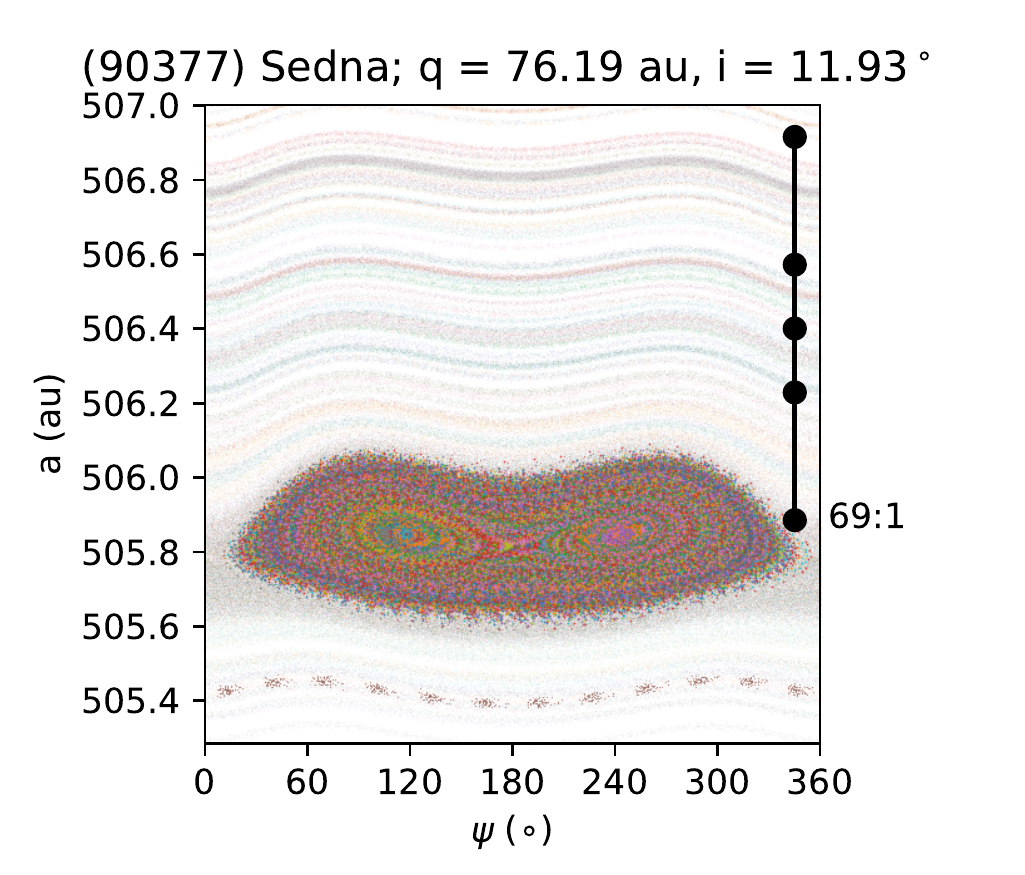} & \hspace{-15pt}
        \includegraphics[width=2.29in]{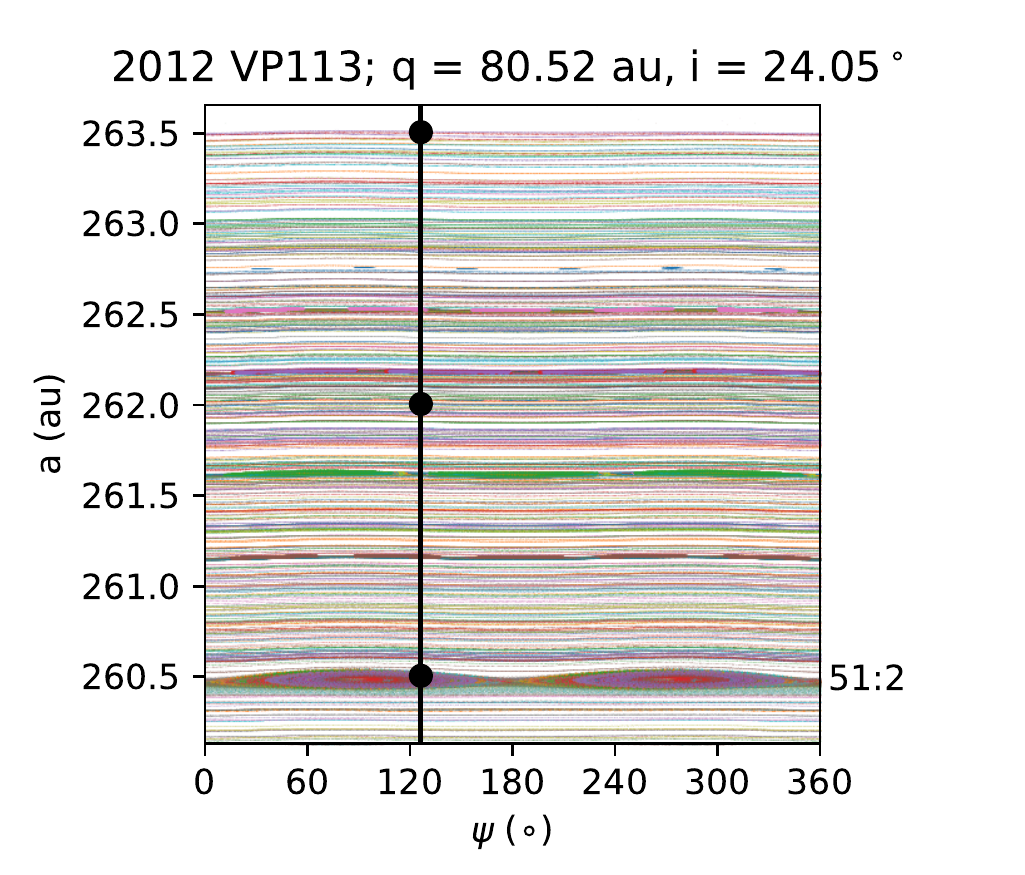} & \hspace{-15pt} \\
    \end{tabular}
    \caption{\textbf{(continued)} The first six panels on this page are sorted from weakest to strongest resonances, which roughly correlates with increasing $q$ (though not strictly increasing $q$ due to the wide range of semimajor axes). 
    Note that for for 2010 GB174, only N:1, N:2, and N:3 resonances within the object's 1$\sigma$ uncertainty range were simulated, and for 2015 RX245 and 2016 SD106, only N:1 and N:2 resonances were simulated; this is due to the large orbital uncertainties these objects span and the diminishing stability seen for the higher order resonances in these regions. A representative subset of the simulation data underlying these figures is available on GitHub (\url{https://github.com/katvolk/TNO-res-maps}).}
\end{figure*}

Two of the objects near our $q=38$~au cutoff for investigation, 2015 GT50 and 2013 SL102, seem likely to be part of the weakly scattering TNO population. 
The non-resonant test particles in these maps show significant $a$ mobility, and the resonant islands are small.
The N:1 resonances in both these maps have lost their symmetric librators, leaving behind only small portions of relatively stable asymmetric islands; we note that this loss of the symmetric libration islands at larger $a$ is consistent with expectations from \cite{Pan:2004}'s analytical model for nearly planet-crossing N:1 populations.
No resonant islands remain for the N:3 and higher-order resonances in the vicinity of these two objects.
Neither of these TNOs have $\psi$ consistent with the remaining stable resonant zones, though they might experience some resonance sticking as they could be near the chaotic edges of those MMRs.
The more distant 2013 RA109 is perhaps in a similar regime. 
Its larger $q$ gives the test particles slightly less mobility in $a$, but the resonant islands are still relatively small, and 2013 RA109's $\psi$ only marginally overlaps the stable islands.

The remaining objects are all in $a$-$q$ regions where Neptune's resonances are strong and have reasonably large stable zones. 
TNO 496315 (2013 GP136) appears firmly outside the stable zone of the 11:1 resonance based on its semimajor axis, though it resides right on the edge of this strong resonance.
TNO 506479 (2003 HB57) is similarly just outside 12:1, though it is more clearly separated from the resonance's chaotic zone; its semimajor axis uncertainty overlaps with the nearby 37:3, but the observed $\psi$ rules out libration in that resonance.
TNOs 505478 (2013 UT15), 2015 RX245, and 2016 SD106 all have semimajor axes that span many strong MMRs but their values of $\psi$ rule out current resonance occupation.
TNO 148209 (2000 CR105) has a best-fit $a$ and $\psi$ consistent with occupation of the symmetric island of Neptune's 20:1 resonance.
The remaining TNOs, (2015 KH163, 2005 RH52, 2018 AD39, 2016 QV89, 2003 SS422, 2014 WB556, 2013 FT28, 474640 (2004 VN112), 2010 GB174, 2016 SA59, and 2015 UN105) all have $a$ and $\psi$ that could be consistent with current occupation of Neptune's MMRs; these objects all require significantly reduced orbital uncertainties to be sure about their resonant status. 

\section{Discussion}\label{s:discussion}

\subsection{The distant reach of Neptune's resonances}\label{ss:disc-res}

Our analysis demonstrates that the vast majority of the TNOs in Table~\ref{t:objs} are not, in fact, beyond the reach of Neptune's dynamical influence via mean motion resonances. 
Simple cuts in semimajor axis and/or perihelion distance are not reliable for determining whether an object's orbit is affected by the known giant planets. 
Furthermore, in the observed sample, attempts to isolate more dynamically `detached' objects by choosing an increasingly large minimum perihelion distance cut can backfire and yield TNOs that are actually in regions dense with Neptune's resonances; see, for example, the additional resonances near 2010 GB174 in Figure~\ref{f:sos-array1} compared to the slightly lower $q$ object 2016 SD106 at the same $a$ range.
This has implications for studies that use the orbital distribution of high-$a$ and high-$q$ objects to infer the presence of additional planets in the outer solar system.
For example, \cite{Brown:2021} use a sample of 11 TNOs from a more restricted $a$ and $q$ range (in addition to considering dynamical stability) to constrain the orbit of their hypothetical Planet Nine; 
two of their TNO sample are excluded from our study for having very poorly constrained orbits, but we have examined the dynamical regimes of the other 9. 
Based on Figure~\ref{f:sos-array1}, Neptune's resonances are still present in the vicinity of all 9 of those TNOs (though extremely weakly in the cases of Sedna and 2012 VP113). 
One third of the \cite{Brown:2021} TNO sample actually have best-fit orbits consistent with being currently resonant with Neptune, and the TNOs not likely to be currently resonant with Neptune are still dynamically affected by 
resonances in their orbital proximity.
 
\begin{figure}
    \centering
    \includegraphics[width=3in]{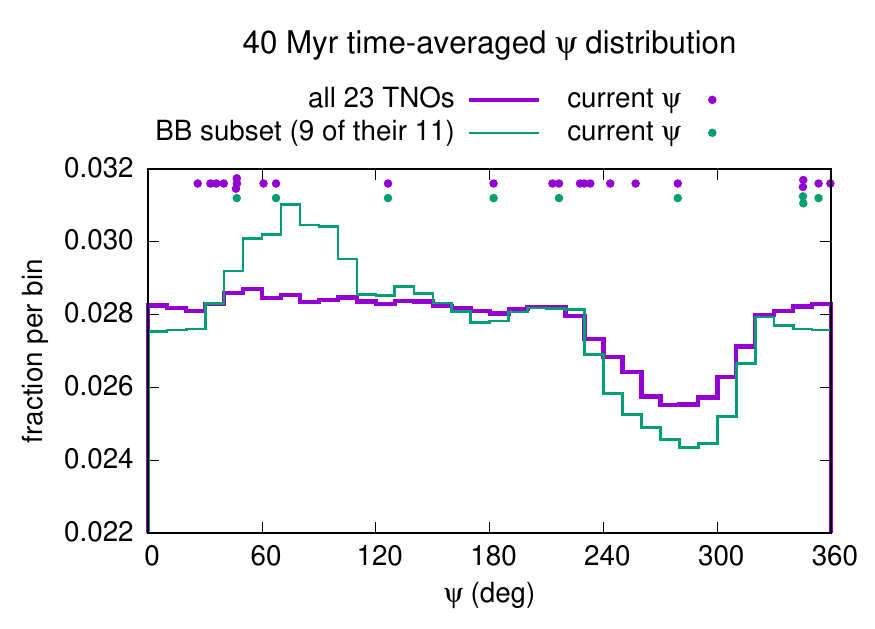}
    \caption{Time averaged $\psi$ distribution for clones of the 23 observed TNOs listed in Table~\ref{t:objs} (thick purple histogram) and the subset of those TNOs that overlap with the \cite{Brown:2021} TNO sample (9 of their 11 total TNOs; green histogram) from a 40 Myr integration of clones sampling each objects' orbit-fit uncertainty; each TNO's current, observed $\psi$ value is shown as a dot near the top of the plot (with arbitrary shifts in y-position to distinguish objects with similar $\psi$). 
    We would expect the histograms to be flat if Neptune's resonances did not have a significant influence on the evolution of the simulated orbits.
    The notable features near $\sim90^\circ$ and $\sim270^\circ$ are due to clones of the TNOs interacting with the N:1 and N:2 resonant libration centers near those values.
    }
    \label{f:psi-dist}
\end{figure}

Resonant (and near-resonant) interactions influence the distribution of TNOs' perihelion locations relative to Neptune, i.e. the distribution of $\psi$. 
To demonstrate this, we examine a time-averaged $\psi$ distribution for the population of TNOs in our sample.
To do this, we integrated clones of all 23 TNOs in Table~\ref{t:objs} for 40 Myr under the influence of the Sun and four giant planets, recording their $\psi$ values at every perihelion passage.
To generate clones of each object, we sampled a Gaussian semimajor axis distribution centered at the best-fit $a$ with $\sigma_a$ given by the JPL orbit uncertainty estimate 300 times. 
For each $a$ value, we then used the observed perihelion distance to assign $e$ and kept all the other orbital elements fixed to their best-fit values. 
This is a simpler scheme than using the full orbit-fit covariance matrix, but provides a good sampling of the available resonant phase space.
The 40 Myr integration timescale is many times longer than any of the libration or circulation timescales for the resonances near the TNOs, thus allowing the clones to fully explore the range of $\psi$ dynamically available.
An object completely unaffected by resonances would have a flat time-averaged distribution of $\psi$ over these integrations.
We determined the $\psi$ distribution for each individual observed TNO and combined them (with equal weighting for each object) into one time-averaged $\psi$ as shown in Figure~\ref{f:psi-dist}.

Figure~\ref{f:psi-dist} shows that the expected $\psi$ of the observed objects is not flat, reflecting the prevalence of clones interacting with Neptune's resonances. 
The notable dip in the distribution near $\psi\approx270$ in particular is due to libration in the trailing asymmetric island of Neptune's N:1 resonances; objects in moderate-amplitude libration tend to spend more time at the extremes of libration than at the center, creating a dip in the time-averaged distribution near the resonance center.
We note that if we consider just the 9 TNOs in our sample that overlap with the \cite{Brown:2021} sample, the non-uniformity of the $\psi$ distribution is enhanced due to the higher prevalence of resonant orbits; the peak near $\psi\approx90^\circ$ is due to the small-amplitude libration of many clones of 474640 (2004 VN112) in the leading asymmetric island of Neptune's 36:1 resonance.
Given the large semimajor axis uncertainties of many of the high-$a$ TNOs, Figure~\ref{f:psi-dist} could change significantly as orbits are refined.
If improved orbit fits converge on the resonant semimajor axis values, it will become less uniform (and vice versa).
We also note that the dynamics of these regions could change in the presence of a currently unseen distant massive planet; see, e.g., \citealt{Hadden:2018,Li:2018} for explorations of the combined dynamics of Neptune and the hypothetical Planet Nine.
\cite{Clement:2021} recently examined how an additional large planet can erode the stability of Neptune's N:1 resonances in the $a\approx150-200$~au region.
As they suggest, the actual prevalence and distribution of distant TNOs near and in resonance with Neptune will provide an important constraint on the presence of additional perturbers.
As we can see from Figure~\ref{f:sos-array1}, orbit-fit improvements are needed for this determination.

\subsection{Implications for the apparent apsidal clustering of distant TNOs}\label{ss:disc-obs}

A population of TNOs that have a non-uniform $\psi$ distribution will also display an apparently non-uniform distribution of longitudes of perihelion ($\varpi$) when sampling only those TNOs that are \textit{currently} near perihelion.
This is because resonant TNOs have constrained values of $\psi=\lambda-\lambda_N$.
If we are considering a specific observational epoch for which Neptune's current location along its orbit is fixed at mean longitude $\lambda_{N0}$, TNOs near perihelion have $\lambda\approx\varpi$ and so by definition must have $\varpi\approx\psi+\lambda_{N0}$.
At the current epoch, $\lambda_{N0} = 354^\circ$, so observable TNOs have $\varpi\approx\psi$.
Thus, because all brightness-limited surveys for larger-$a$ TNOs are strongly biased toward TNOs that are at or near perihelion, the \textit{currently observable} distribution of $\varpi$ will be non-uniform if the TNO population being observed is significantly affected by resonances. 
It is important to emphasize that this need not reflect an \textit{intrinsic} non-uniformity in the $\varpi$ distribution; it is instead an \textit{epoch-dependent} non-uniformity.
If one were to conduct an all-sky survey of a fully resonant TNO population over a full orbital period for Neptune (164 years), one would have no preference toward a particular range of $\varpi$ and could recover the intrinsic $\varpi$ distribution directly, even from a perihelion-biased sample; but because TNO surveys have a time baseline much shorter than Neptune's orbital period, even all-sky surveys cannot necessarily probe the full range of their $\varpi$.
This `bias' in the currently observable $\varpi$ distribution for resonant populations is in addition to any biases in the observed TNO population that result from the non-uniform sky coverage of TNO surveys to date. 
See \cite{Shankman:2017}, \cite{Napier:2021}, and \cite{Bernardinelli:2022} for examples of how pointing biases in the surveys that have discovered significant portions of the known TNOs result in very clear non-uniform observed $\varpi$ distributions.
These works found that the observed clustering in $\varpi$ for the subset of high-$a$ TNOs with well-characterized discovery circumstances is consistent with the survey biases.
They did not, however, consider any additional effects from dynamically induced non-uniformity in the distribution of $\varpi$ for TNOs currently near perihelion.
This would not change their conclusions as the pointing biases are much larger in magnitude than resonant-induced epoch-dependent biases, but we will estimate the size of the latter effect here for demonstration purposes.

\begin{figure}
    \centering
    \includegraphics[width=3in]{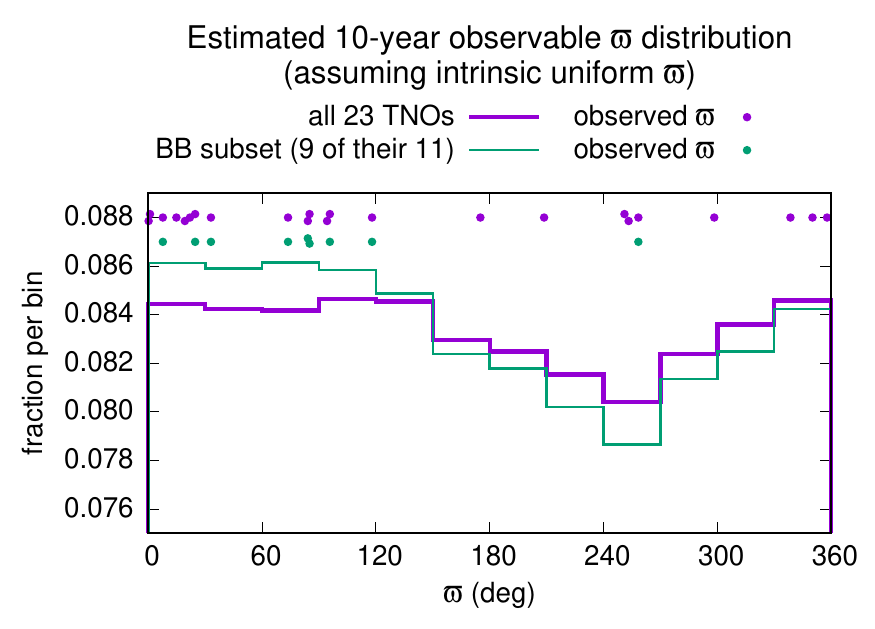}
    \caption{Estimated observable $\varpi$ distribution for clones of the 23 observed TNOs listed in Table~\ref{t:objs} (thick purple histogram) and the subset of those TNOs that overlap with the \cite{Brown:2021} TNO sample (9 of their 11 total TNOs; green histogram) assuming that the true, intrinsic $\varpi$ is uniform and the population has the time-averaged $\psi$ distribution relative to Neptune as shown in Figure~\ref{f:psi-dist}; each TNO's actual observed $\varpi$ value is shown as a dot near the top of the plot (with arbitrary shifts in y-position to distinguish objects with similar $\varpi$). 
    We assume an object is `observable' if at any point over a 10-year timespan its estimated apparent brightness is 24.5 magnitudes or brighter. See Section~\ref{ss:disc-obs} for details.
    }
    \label{f:varpi-dist}
\end{figure}

Given the time-averaged expected distribution of $\psi$ for the 23 TNOs in Table~\ref{t:objs} (see Figure~\ref{f:psi-dist}), we can make a rough estimate of the dynamically induced `bias' in the current observable $\varpi$ distribution under the assumption that the intrinsic $\varpi$ is actually uniform.
For the purposes of this illustration, we will consider an object `observable' if it has an estimated apparent magnitude brighter than 24.5 at any point over a 10-year span starting at the current epoch, regardless of position in the sky; this is a \textit{very} crude approximation of a deep, long-baseline, large-area survey like the Vera Rubin Observatory's upcoming Legacy Survey of Space and Time (LSST; see, e.g., \citealt{Ivezic:2019}).
To do this, we generate a large synthetic population representative of each TNO.
We use the integrations described above (Section~\ref{ss:disc-res}) to determine the distribution of $a$, $e$, $i$, and $\psi$ for each object, assign uniform random distributions of $\varpi$ and $\Omega$, and assume a simple power law for the intrinsic brightness distribution with a slope of 0.9 from $H=1.5-9$ (the bright-end distribution for the scattering TNO population; see, e.g., \citealt{Lawler:2018}); the mean anomaly of the synthetic object is then determined to be $M= \psi + \lambda_N - \varpi$.
Then, assuming observations take place at opposition, we can convert this starting position and $H$ magnitude to an estimated brightest apparent magnitude over a 10 year period by propagating $M$ forward in time.
We repeat this process to build up a catalog of synthetic detectable objects based on each real TNO to determine the expected observable $\varpi$ distribution based on its $\psi$ distribution. 
These distributions are then combined to produce Figure~\ref{f:varpi-dist}, which shows that the resonant interactions do indeed produce a slightly non-uniform expected observable $\varpi$ distribution for the known high-$a$ TNOs.

For the full sample, there is a $\sim5\%$ variation in the expected number of TNOs from the most to least probable $\varpi$ values; for the 9 TNOs overlapping with the \cite{Brown:2021} sample, there is a $\sim9\%$ variation.
We note that a complete understanding of this dynamical bias requires a much better understanding of and model for the intrinsic $a$ and $q$ distribution for the distant TNOs, which is beyond the scope of this work. 
The model presented here is meant only to be illustrative.
This dynamically-induced `bias' is clearly not large enough to explain the non-uniform observed distribution of $\varpi$ for these TNOs; when comparing the ranges $\varpi=0-180^\circ$ and $\varpi=180-360^\circ$, it only results in an expected 1-2\% asymmetry in the distribution of the current observed objects. 
However, the dynamical bias for the current set of TNOs is toward the cluster in observed $\varpi$ and could thus add to the pointing biases that already provide a promising explanation for the non-uniformity \citep{Shankman:2017,Napier:2021,Bernardinelli:2022}.

For completeness, we also investigate whether we should expect the \textit{intrinsic} distribution of $\varpi$ for a high-$a$ TNO population to be uniform if they are strongly affected by resonances.  
We did this by re-examining existing integrations of a model of the closer-in scattering population of TNOs, which is known to be significantly affected by the phenomenon of resonance sticking \citep[e.g.][]{Duncan:1997,Lykawka:2007}.
\cite{Yu:2018} integrated the \cite{Kaib:2011} model of the scattering TNO population for 1 Gyr, finding that the 40\% of the $a=30-100$~au population was temporarily resonant with Neptune at any given time.
The initial $\varpi$ distribution for that simulation was uniform, so we examined the final snapshot at 1 Gyr to look for any changes that might have resulted from those resonant interactions.
We find no statistically significant evidence of the resonances generating a non-uniform $\varpi$ distribution, either for the $a=30-100$~au or for the full semimajor axis range of the modeled scattering population (for which we do not have a quantified resonant fraction from \citealt{Yu:2018}).
While this population is not a perfect analog for the high-$a$, high-$q$ TNO populations of interest here, it is a population known to be heavily influenced by Neptune's resonances and it does not show evidence of a sculpted $\varpi$ distribution. 
This might merit future investigation for the most distant TNOs, whose orbital precession rates are very slow and thus perhaps more likely to retain dynamical signals in their $\varpi$ distributions (see, e.g., discussion in \citealt{Clement:2020}).

\subsection{Future work}\label{ss:future}

In addition to investigating the dynamical regimes of known TNOs, our new mapping approach provides a way to more generally probe for the importance of Neptune's resonances and how the other giant planets influence the structure of Neptune's resonances in different TNO orbital regions (see Appendix~\ref{appendix}). 
We are exploring these questions more broadly and expect to obtain new insights into the dynamical distribution of TNOs,  when combined with the expected large increase TNO discoveries when LSST is underway. 
Understanding the full range of possible  dynamical behaviors of TNOs will help identify which features of the observed dynamical distribution of the TNOs are signatures of planet migration, prior perturbers, or unseen present-day perturbers.

We also plan to expand this mapping approach to a wider range of observed TNOs.
For each individual TNO considered in this work, the mapping simulations described in Section~\ref{ss:methods} typically required a few hundred CPU hours to complete.
This is more CPU time than typical, $\sim$10~Myr integrations of a comparable number of test particles sampled from an observed TNO's orbit, partly due to the requirement that each perihelion passage be very well-resolved to record the parameters needed for our maps and partly due to our choice of integrator. 
However, as noted above, the maps need only be computed once as they will remain valid even as a TNO's orbit is refined over time by additional observations.
We also note that we made no effort to optimize our mapping approach for computational efficiency because it was not necessary considering the small number of TNOs in this work.
As we discover more of these distant TNOs or expand the analysis to closer-in objects, the CPU time requirements per object could become limiting.
Several steps could be taken to improve efficiency in our mapping simulations. 
The most fruitful would likely be to switch to the \textsc{whfast} integrator \citep{Rein:whfast}, which is significantly faster than \textsc{ias15}, and add an interpolation step near a test particle's perihelion passage to estimate the state vector at perihelion rather than relying on an integration timestep small enough to directly resolve it. 
We used \textsc{ias15} in this study for its very high accuracy, but \textsc{whfast} would work very well for the TNOs of interest here because their large perihelion distances mean they don't experience close encounters with the planets (which \textsc{whfast} cannot accurately resolve).
We also note that the resonances could be resolved with sparser sampling in $\psi$ than we have used here, again reducing the CPU time required to produce the maps.
We leave such modifications to our procedure for future work.

\section{Summary and Conclusions}\label{s:summary}

We have examined the dynamical regimes of 23 TNOs with semimajor axes larger than 150~au and perihelion distances larger than 38~au. 
Nearly half of this sample (11 objects) have observed $a$ and $\psi$ values that are consistent with stable libration zones of Neptune's external resonances. Of particular note amongst these TNOs is that the best-fit orbits of 148209 (2000 CR105) and 474640 (2004 VN112) fall within the stable regions of Neptune's 20:1 and 36:1 resonances, respectively.
Two objects (Sedna and 2012 VP113) continue to stand out as being on orbits that are relatively unaffected by the giant planets, though Neptune's resonances are still weakly present even at such extreme $a,q$ combinations.
One object (2015 KG163) likely belongs to the weakly bound, high-$a$, high-$q$ diffusing class of TNOs, with no resonances in its vicinity. 
Two or three objects (2015 GT50, 2013 SL102, and possibly 2013 RA109) are in the regime of moderate to weak scattering, with $\psi$ marginally consistent with the possibility of temporary sticking to the chaotic boundaries of Neptune's N:1 MMRs. 
Only five objects (besides Sedna and 2012 VP113) can definitively be said to reside on relatively stable orbits that do not currently overlap with the stable regions of strong resonances (506479, 496315, 505478, 2016 SD106, and 2015 RX245); this determination relies mostly on $\psi$ rather than $a$ as all five objects are in regions that have strong resonances within their semimajor axis uncertainty ranges. 
This highlights the usefulness of our dynamical mapping approach when orbit-fit uncertainties are large and when the uncertainties themselves are poorly determined. 
Secure dynamical classification of TNOs is only possible when the orbit-fit uncertainties are well-understood. 
We strongly suggest that the orbital parameter uncertainties be made readily available in all public databases of solar system small body orbits.
Follow-up observations of the majority of the distant TNOs are required to securely determine their current dynamical state.

Using this sample of TNOs, we have demonstrated that Neptune's resonances extend out significantly further than often assumed and that simple cuts in $a$ and $q$ cannot reliably be used to determine whether an object is dynamically isolated from the known giant planets.
We showed that the current observed sample of high-$a$ and high-$q$ TNOs maintains a non-uniform distribution of perihelion locations relative to Neptune ($\psi$ values) due to the prevalence of resonant interactions. 
Even for a population with an intrinsically uniform distribution of perihelion longitudes ($\varpi$), the non-uniform $\psi$ distribution from resonant interactions can lead to an apparent bias in $\varpi$ values for TNOs currently near their perihelia and thus bright enough to be detectable; this bias will only disappear when the time baseline for TNO discoveries is comparable to Neptune's orbital period.
While the exact magnitude of this effect on even all-sky observations of the distant TNO population is difficult to quantify because our models of the $a$ and $q$ distribution of this population are not well-constrained, we estimate that it results in a few percent asymmetry in the expected observed $\varpi$ distribution.
This is not a large enough bias to explain the much larger observed non-uniformity in $\varpi$, which is likely better explained by pointing biases in discovery surveys \citep{Shankman:2017,Napier:2021,Bernardinelli:2022}, but it is an additive effect.
It is also a bias that is not remedied by performing an all-sky survey and thus should be kept in mind when analyzing future, much larger sets of observed distant TNOs.

\vspace{12pt}
\noindent{\it Acknowledgements:}
We thank Matt Clement for a helpful review of the manuscript.
This work was supported by grants from NSF (AST-1824869) and NASA (80NSSC19K0785).
This work used High Performance Computing (HPC) resources supported by the University of Arizona TRIF, UITS, and Research, Innovation, and Impact (RII) and maintained by the UArizona Research Technologies department.

\facilities{ADS}
\software{rebound}

\appendix
\section{Additional validation of the modified N-body Poincar\`e map}\label{appendix}

\begin{figure}
    \centering
    \begin{tabular}{c c}
    \hspace{-10pt} 
    \includegraphics[width=3.85in]{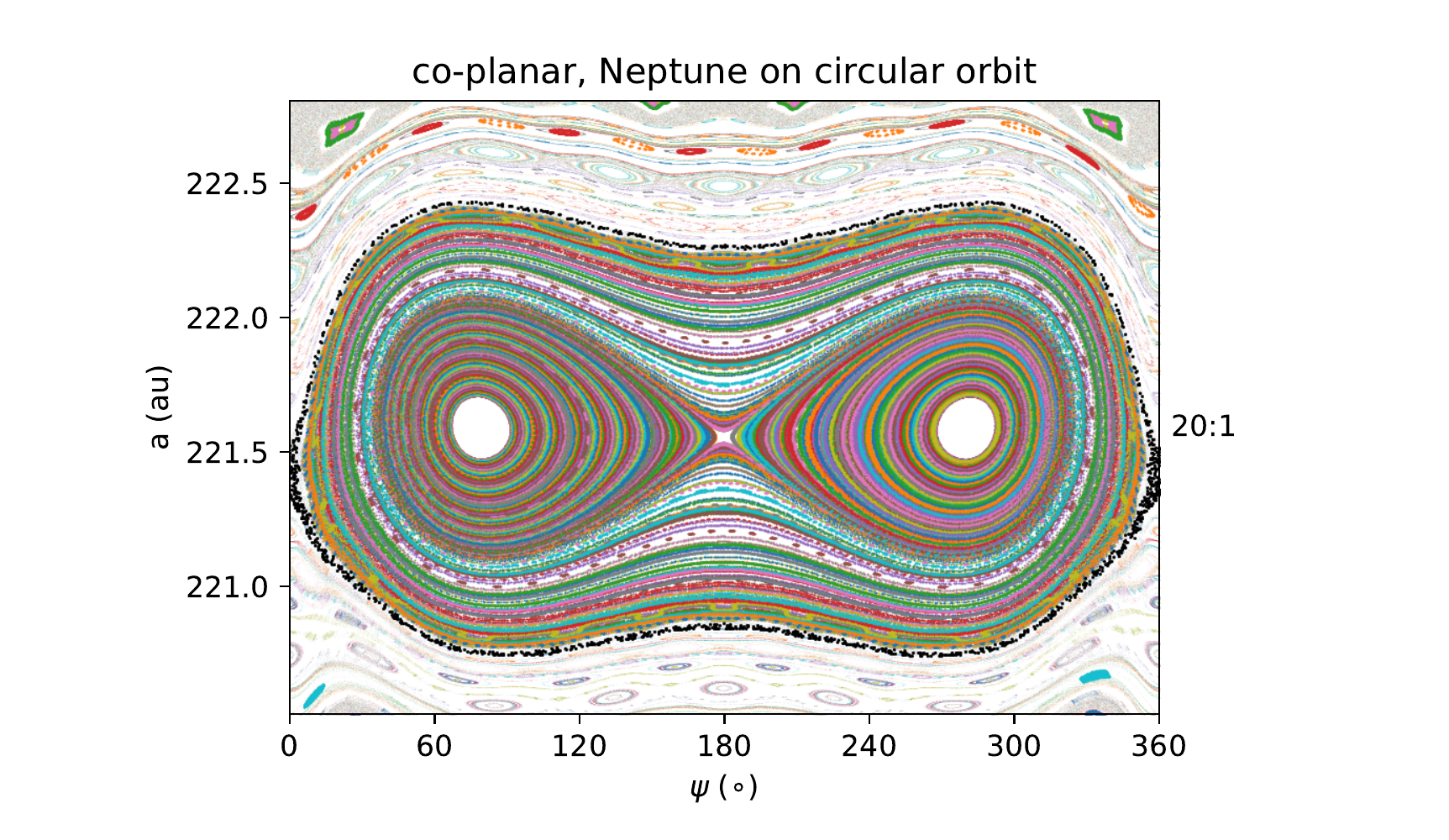} & 
    \hspace{-25pt}
    \includegraphics[width=3.85in]{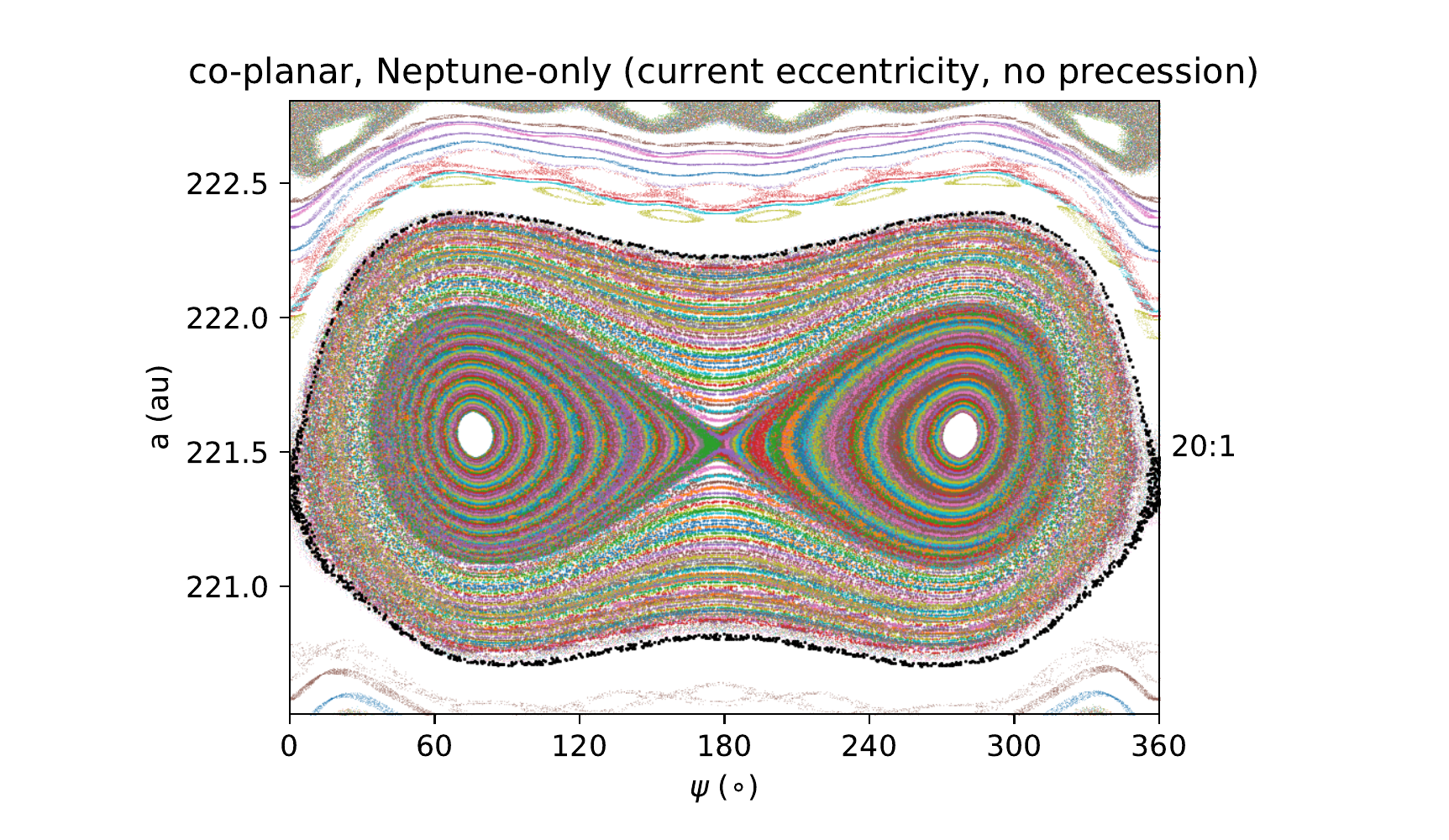}\\
    \hspace{-10pt} 
    \includegraphics[width=3.85in]{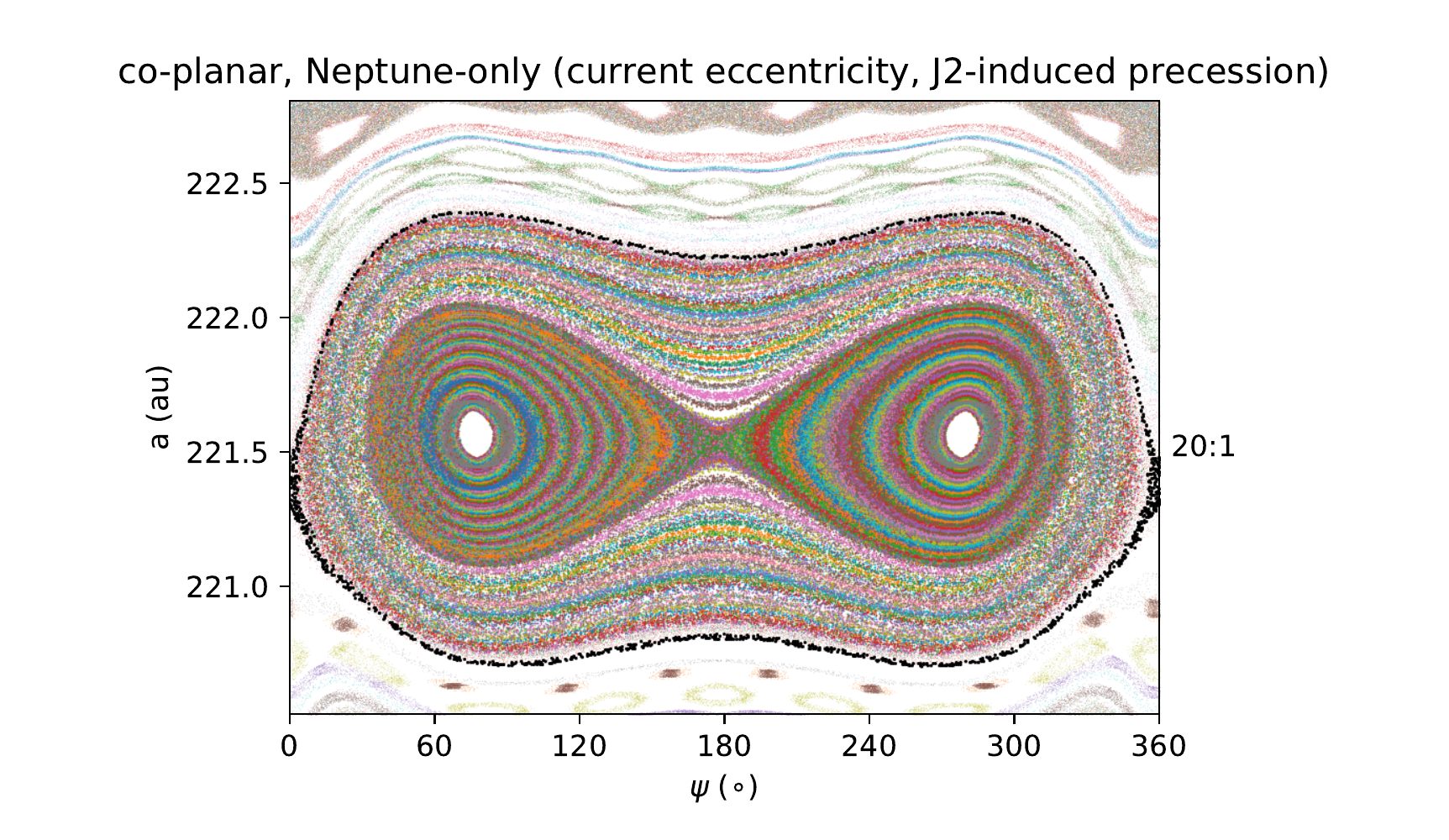} & 
    \hspace{-25pt}
        \includegraphics[width=3.85in]{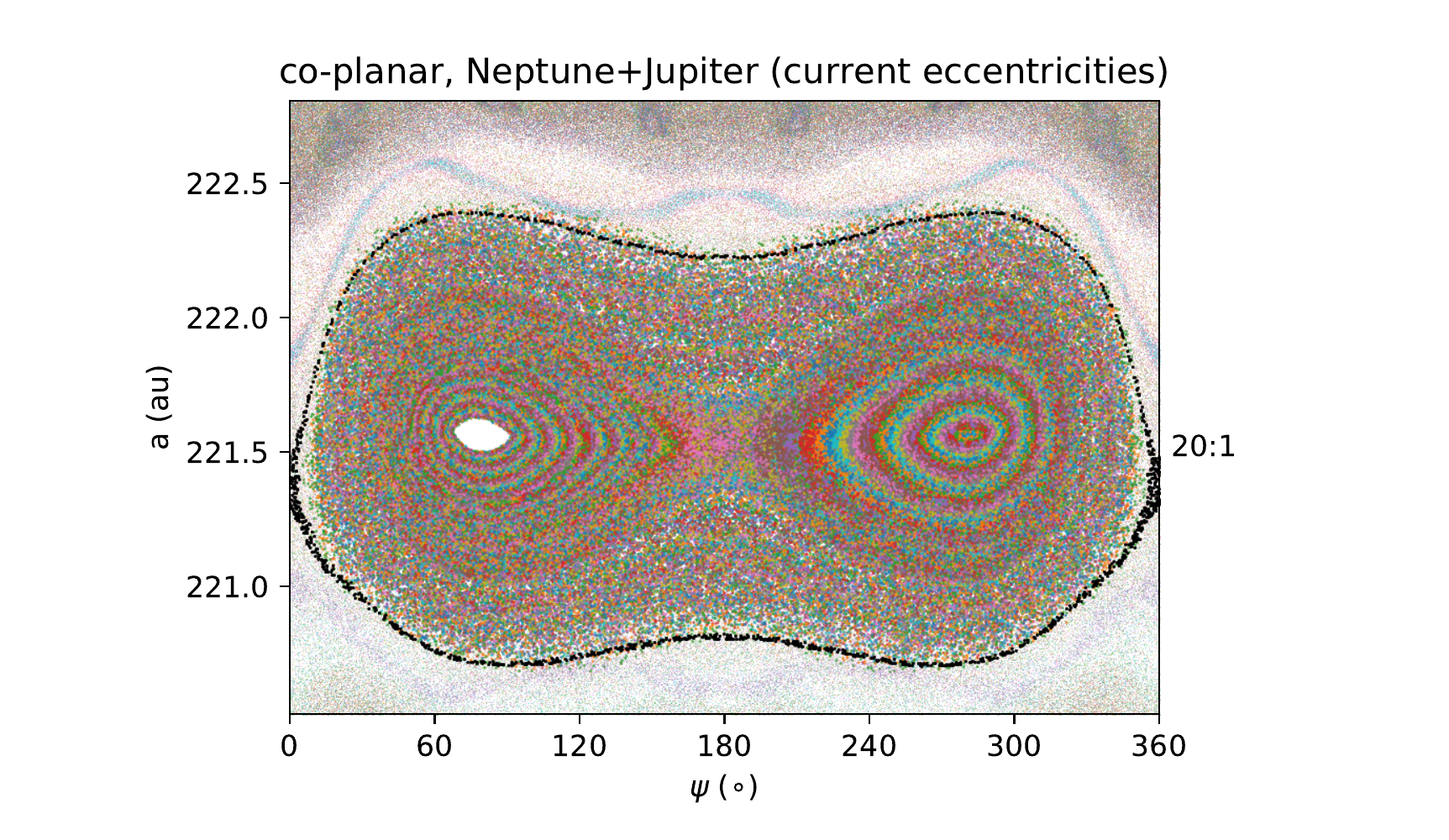}\\
   \hspace{-10pt}
    \includegraphics[width=3.85in]{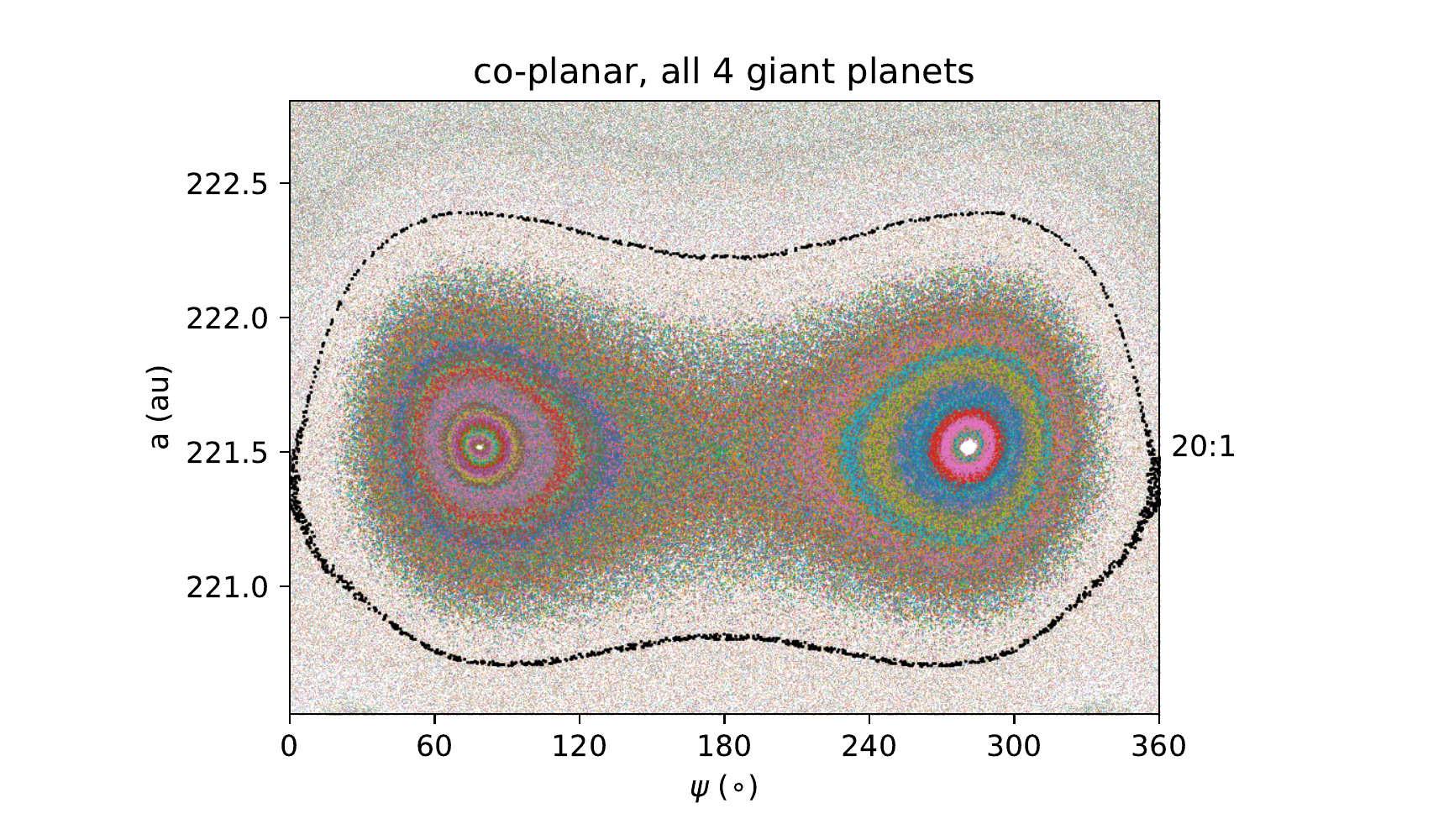} &
     \hspace{-25pt}
        \includegraphics[width=3.85in]{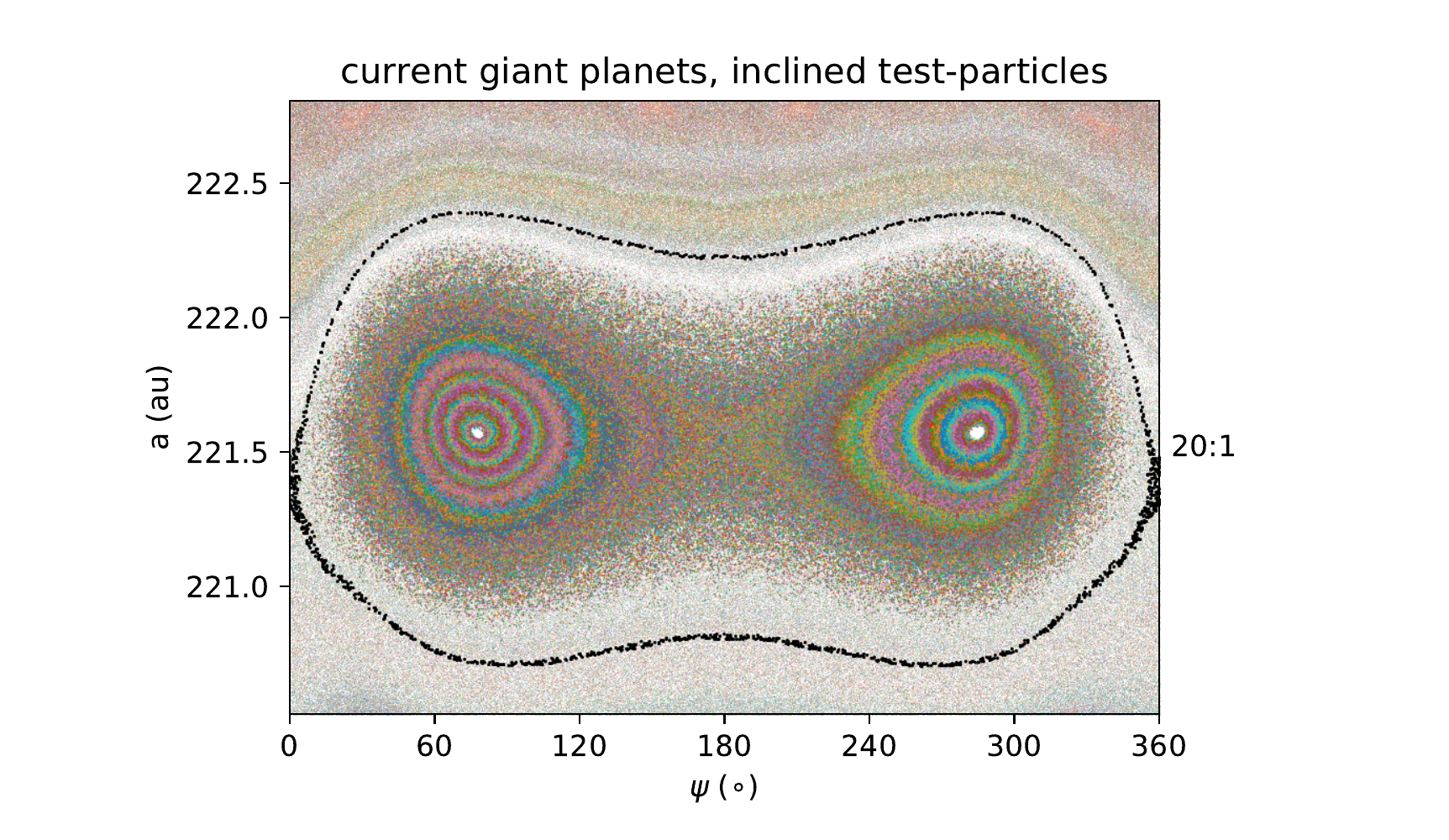}  \\
        \end{tabular}
    \caption{Poincar\`e maps of Neptune's 20:1 MMR (for $q=44$~au) in the circular, planar, restricted three body problem (top left) compared to simulations with increasing complexity. 
    In the top right, test particles evolve on orbits under the influence of the Sun plus a co-planar Neptune with an eccentricity of 0.02 (and a fixed orbital orientation); this slightly increases the `fuzziness' of the particle paths in the $a$-$\psi$ plane. 
    In the middle left, we have added perihelion precession to Neptune's orbit using a J2 term; this again slightly increases the fuzziness. 
    In the middle right, we have added Jupiter to the system at its current eccentricity (both planets and the test particles remain co-planar); this significantly blurs the distinction between separate particle paths as Jupiter causes Neptune's semimajor axis and eccentricity to vary over time. 
    In the bottom left, we include all four giant planets (at their current eccentricities) on co-planar orbits with the each other and the test particles; the perturbations from the additional planets reduce the extent of the stable symmetric libration zone. 
    In the bottom right panel, the giant planets are all on their current, mutually-inclined orbits, and the test particles are initialized on orbits inclined by $\sim$23$^\circ$ (inclination and node matching TNO 148209).}
    \label{fig:map-evolution}
\end{figure}

As noted in Section~\ref{ss:methods}, the Poincar\`e maps presented here are modified from the traditional  Poincar\`e maps produced in the planar circular restricted three body problem. 
One modification we make, even in our maps of test particles in the circular restricted three body problem (left panel of Figure~\ref{f:ex}; top left panel of Figure~\ref{fig:map-evolution}) is that we initialize our test particles with varying $a$ and $\psi$ at constant perihelion distance $q$ rather than with with varying $a$ and $\psi$ at constant values of the Jacobi integral (as was done in \citealt{Wang:2017,Lan:2019}). 
Our motivation for fixing $q$ is that for observed TNOs, $q$ is typically one of the better-determined orbital elements. 
Additionally, for high-eccentricity orbits, constant $q$ is approximately equivalent to constant Jacobi integral. 
In the restricted three body problem, the Jacobi integral can be approximately expressed as the Tisserand parameter,
\begin{equation}
    T = \frac{1}{2a} + \sqrt{a(1-e^2)}\cos{i}
\end{equation}
(see discussion in \citealt{Wang:2017}).
As $a$ is varied amongst the test particle initial conditions for a single map, $T$ can be used to calculate the value of $e$ that should be assigned to each particle to maintain a constant Jacobi integral.
For orbits with eccentricities typical of the TNOs in Table~\ref{t:objs} ($e\gtrsim0.7$), instead assigning $e$ based on maintaining constant $q$ results in eccentricities that differ from those calculated from $T$ by less than 0.2\% across a $\sim10$~au range. 
Thus, maintaining constant $q$ (to match the observed value) rather than constant Jacobi integral across a typical semimajor axis uncertainty range is a reasonable modification to the Poincar\`e mapping calculations for the restricted three body problem.
Additionally, when the model is expanded to include the perturbations of the other giant planets, the Jacobi integral is no longer a conserved quantity, further justifying the modification.
 
We performed a number of tests to explore how the modified Poincar\`e maps described in Section~\ref{ss:methods} evolve as complexity is added to our simulations.
Figure~\ref{fig:map-evolution} shows how our 2D projection of the 20:1 resonance in the ($a$,$\psi$) plane changes compared to the simplest case (the circular planar restricted three body model of the Sun--Neptune--test particle) as additional perturbations and/or perturbers are added to the model.
We start by allowing Neptune to have an eccentric orbit ($e=0.02$) with a fixed orbital orientation. 
This slightly increases the `fuzziness' of the particle paths in the $a$-$\psi$ plane (top right panel) compared to the simplest problem, and some of the very high-order resonant chains surrounding the 20:1 become less clear.
We next added perihelion precession to Neptune's orbit by adding a J2 term to our integrations (we use the J2 terms for the other giant planets from \citealt{Malhotra:2022}).
This adds additional fuzziness to the closed curves representing stable resonant libration, though does not eliminate the higher-order islands (middle left panel). 
Next, we added Jupiter to the integrated system at its current eccentricity, but kept both planets and the test particles in the same plane.
This significantly reduces the separation between the libration paths of individual particles (middle right panel) and washes out the nearby higher-order resonant islands, likely because Neptune's orbit experiences more significant time-varying changes;
in addition to perihelion direction precession, Neptune's semimajor axis and eccentricity vary over time due to the perturbations from Jupiter. 
In the final two models, we include all four giant planets at their current eccentricities. 
In one model, we keep them all on co-planar orbits with each other and the test particles (lower left panel) and in the other we allow their orbits to assume their current inclinations and place the test particles on inclined orbits (lower right panel; the same inclination as in Figure~\ref{f:ex}). 
In both cases, the extent of the stable symmetric libration zone for the 20:1 resonance is reduced compared to the simpler models.
This is because Uranus has a relatively strong effect on Neptune's orbital evolution through both secular perturbations and also because Neptune and Uranus have orbital periods that are themselves close to a 2:1 resonant ratio. 
In future work we will explore in more detail how Uranus may influence Neptune's resonances, but, for the present work, we have shown in this Appendix that including all four giant planets is important for fully exploring the influence of Neptune's resonances on observed TNOs, and that our modified Poincar\`e maps are able to capture and visualize the relevant dynamics in simple 2D plots.

\end{document}